\documentclass[fleqn,usenatbib]{mnras}
\usepackage{newtxtext,newtxmath}

\usepackage[T1]{fontenc}
\usepackage{ae,aecompl}
\usepackage{graphicx}	% Including figure files
\usepackage{amsmath}	% Advanced maths commands
\usepackage{ulem}
\usepackage{multirow}

\newcommand{\UM}{\textsc{UniverseMachine}}

\newcommand{\appropto}{\mathrel{\vcenter{
  \offinterlineskip\halign{\hfil$##$\cr
    \propto\cr\noalign{\kern2pt}\sim\cr\noalign{\kern-2pt}}}}}

\title[Observing DM-Galaxy Growth Correlations]{Observing Correlations Between Dark Matter Accretion and Galaxy Growth: I. Recent Star Formation Activity in Isolated Milky Way-Mass Galaxies}

\author[C. O'Donnell et al.]{
Christine O'Donnell$^{1}$\thanks{E-mail: Christine.ODonnell@asu.edu (CO)},
Peter Behroozi$^{2}$,
Surhud More$^{3,4}$,
\\
$^{1}$Center for Gender Equity in Science \& Technology, Arizona State University, Tempe, AZ 85281, USA\\
$^{2}$Department of Astronomy and Steward Observatory, University of Arizona, Tucson, AZ 85721, USA\\
$^{3}$Inter-University Centre for Astronomy and Astrophysics, Post bag 4, Ganeshkhind, Pune 411007, India\\
$^{4}$Kavli Institute for the Physics and Mathematics of the Universe, 5-1-5 Kashiwanoha, Kashiwa, 2778583, Japan\\
}

\date{Accepted XXX. Received YYY; in original form ZZZ}
\pubyear{2020}

\begin{document}
\label{firstpage}
\pagerange{\pageref{firstpage}--\pageref{lastpage}}
\maketitle

\begin{abstract}
The correlation between fresh gas accretion onto haloes and galaxy star formation is critical to understanding galaxy formation. Different theoretical models have predicted different correlation strengths between halo accretion rates and galaxy star formation rates, ranging from strong positive correlations to little or no correlation. Here, we present a technique to observationally measure this correlation strength for isolated Milky Way-mass galaxies with $z < 0.123$.  This technique is based on correlations between dark matter accretion rates and the projected density profile of neighbouring galaxies; these correlations also underlie past work with splashback radii.  We apply our technique to both observed galaxies in the Sloan Digital Sky Survey as well as simulated galaxies in the \textsc{UniverseMachine} where we can test any desired correlation strength. We find that positive correlations between dark matter accretion and recent star formation activity are ruled out with $\gtrsim 85\%$ confidence. Our results suggest that star formation activity may not be correlated with fresh accretion for isolated Milky Way-mass galaxies at $z=0$ and that other processes, such as gas recycling, dominate further galaxy growth. 
\end{abstract}


\section{Introduction}
\label{sec:intro}

Under the Lambda Cold Dark Matter ($\Lambda$CDM) framework, galaxies formed at the centres of dark matter haloes \citep[see][for reviews]{Somervile15, WT18}. As the Universe evolved, gas was able to dissipate energy and fall to the centres of these haloes. Stars formed if enough gas coalesced, leading to the galaxies we observe today. Given these formation processes, we expect that halo properties should be correlated with galaxy properties. For example, many studies have found a strong correlation between halo mass and stellar mass \citep[e.g.,][]{Leauthaud12,Tinker17b, Behroozi19}.

At large distances, gravity dominates, and so the ratio of infalling gas to infalling dark matter is expected to be the cosmic baryon fraction. If infalling gas also tracks infalling dark matter at smaller scales, then we expect to see a strong positive correlation between dark matter accretion rates and galaxy star formation rates. \cite{WetzelNagai15} found a tight relation between halo accretion and galaxy growth. They found that as a halo accreted material, dark matter was deposited in a shell-like manner at $\gtrsim R_{200\mathrm{m}}(z)$ , consistent with results from \cite{Diemer13} that found little to no halo growth within $\sim R_{200\mathrm{m}}$ from $z=1$ to $z=0$. However, infalling gas decoupled from the dark matter at $\sim 2R_{200\mathrm{m}}$ and continued to accrete to smaller radii. Thus, star formation rates tracked the dark matter accretion rates. Other models have assumed a perfect positive correlation between star formation rates and halo growth rates. For example, \cite{Becker15}, \cite{Rodriguez-Puebla16}, and \cite{Cohn17} assumed models that directly couple halo growth and galaxy star formation rates. \cite{Moster18} presented an empirical model for galaxy formation since $z \sim 10$ and assumed perfect correlation between a central galaxy’s star formation and its halo accretion. 

However, other studies have found little to no correlation between halo accretion rates and star formation rates. \cite{Tinker17} studied star formation in the central galaxies of galaxy groups as a function of local density. They found only a slight increase in the fraction of quenched galaxies for high halo masses ($M_* \gtrsim 10^{10} M_\odot / h^2$) from low to high densities.  Because halo assembly rates vary strongly with local density \citep[e.g.,][]{Lee17}, this implied a weak correlation between halo growth and galaxy assembly. Similarly, \cite{Behroozi15} did not find a correlation between star formation rates and major halo mergers. Further, simulations of massive Milky Way-mass haloes at low redshifts ($z \lesssim 1$) suggested that gas accretion onto haloes was primarily through `hot mode accretion', which was quasi-spherical and less efficient \citep[e.g.,][]{Keres05, Nelson13, Dekel06}, and \cite{Nelson15} found that feedback processes, including radiative cooling, winds, and suppermassive black holes, strongly suppressed the accretion of this gas onto galaxies. These results are consistent with models of galaxy growth where gas recycling happens on short timescales and is responsible for the majority of star formation \citep[e.g.][]{vandeVoort16}. Additionally, \cite{Muratov15} and \cite{vandeVoort16} (and references within) suggest that outflows from processes such as supernovae and active galactic nuclei can prevent gas from accreting onto a central galaxy and leading to star formation. \cite{Muratov15} found that these outflows are most significant at high redshifts, and at lower redshifts, the ejected material forms a reservoir of enriched gas that may be recycled for further star formation. 

\begin{figure}
    \centering
    \includegraphics[width=\columnwidth]{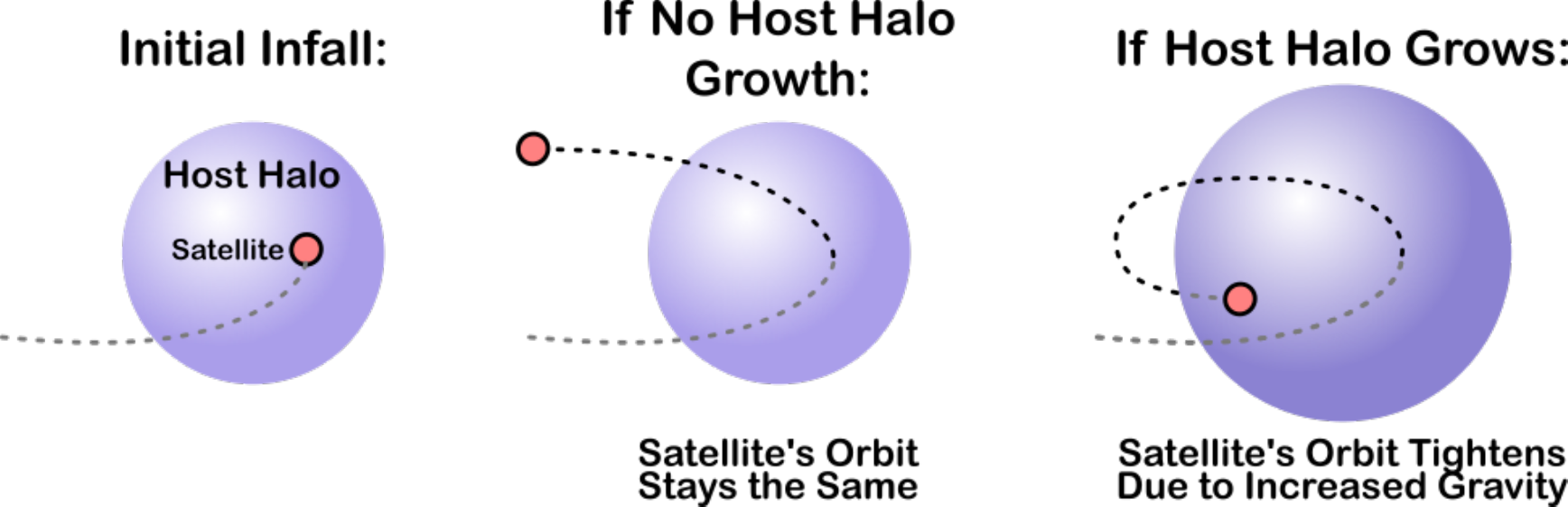}\\[-0.5ex]
    \caption{Schematic of the effect of halo accretion on satellite orbits. As a satellite enters a larger host halo, its orbit is sensitive to changes in the host halo's gravitational potential well. For host haloes that do not grow very much, satellite orbits stay large (middle panel). For host haloes that accrete more material, their gravitational potential wells deepen, tightening the orbits of satellite galaxies (right panel).}
    \label{fig:orbits_schematic}
\end{figure}

To constrain the correlation between halo accretion and star formation rates, observational tests are needed. Our technique builds on past work \citep{Diemer14, More15, More16, Baxter17} to measure the \textit{splashback radius}, the location at which accreted material reaches its first orbital apocentre. As a halo accretes more dark matter, its gravitational potential well deepens, which tightens the orbits of satellite galaxies and steepens the halo density profile (Fig.~\ref{fig:orbits_schematic}). \cite{Diemer14} found that steepening of the halo density profile is stronger for more massive or rapidly accreting haloes.  Similarly, \cite{More15} found that the splashback radius decreases for more rapidly-accreting haloes. \cite{More16} and \cite{Baxter17} developed observational techniques using SDSS photometric data to measure splashback radii.  Their basic technique involved measuring excess photometric galaxy counts (via background subtraction) around target clusters and then using the slope of the radial profile of excess photometric galaxies to identify the splashback radius.  Both studies stacked the radial profiles around thousands of clusters to average halo-to-halo scatter and increase signal-to-noise.   

On cluster scales ($M_\mathrm{halo} \gtrsim 10^{14} M_\odot$), most central galaxies are quiescent \citep{Yang09}.  Studying the correlation between dark matter accretion and galaxy formation requires a technique that works for lower-mass haloes ($M_\mathrm{halo} \sim 10^{12}-10^{13} M_\odot$).  Yet, lower-mass haloes often have larger neighbouring haloes, which contaminate the distributions of nearby galaxies and dark matter with their own orbiting material \citep{More15} and smear out the splashback feature. \cite{Deason20} analysed cosmological simulations of Milky Way-mass haloes and found that if these haloes are isolated, there are clear splashback features. By definition, isolated haloes are the largest halo (and thus the dominant source of gravity) in their nearby environment, and so exhibit much stronger splashback features and orbital changes correlated with accretion even at the lower masses of interest.  

Because lower-mass haloes host fainter galaxies, they are most easily detectable at lower redshifts.  As a result, the haloes' angular sizes on the sky are larger, leading to larger noise levels from background photometric galaxies.  To work around this, we develop a technique that uses the entire radial distribution of nearby photometric galaxies and optimally weights the stacking process to minimise background source contamination.  We use simulated galaxies (from the \textsc{UniverseMachine} empirical model; \citealt{Behroozi19}) to validate that the technique can measure the correlation strength between halo assembly and galaxy assembly.  As a proof-of-concept, we also apply the technique to Milky-Way mass galaxies ($10.5 < \log_{10}(M_*/M_\odot) < 11$) in the Sloan Digital Sky Survey (SDSS) DR16 \citep{SDSS_DR16}.

This paper is organised as follows: In \S\ref{sec:data}, we describe the data sets used in this paper, including observed spectroscopic data to identify isolated Milky Way-mass host galaxies (\S\ref{sec:obs_spec}), observed photometric data to measure neighbour density distributions (\S\ref{sec:obs_phot}), and simulation data to demonstrate our technique for constraining correlations with halo accretion rates (\S\ref{sec:UM}). In \S\ref{sec:methods}, we describe our analysis methods, including the background-subtraction technique to generate neighbour density distributions (\S\ref{sec:methods_overview}), the metric for quantifying the shapes of the neighbour density distributions (\S\ref{sec:shape_ratio}), and how we account for systematics and selection effects (\S\ref{sec:systematics}). We present results in \S\ref{sec:results} and conclude in \S\ref{sec:disc_conc}. Throughout this paper, we adopt a flat $\Lambda$CDM model with $h = 0.677$, $\Omega_M = 0.307$, and $\Omega_{\Lambda} = 0.693$, consistent with \textit{Planck} 2018 results \citep{Planck18}.

\section{Data}
\label{sec:data}

To constrain the correlation between dark matter accretion and specific star formation rates (a measure of recent star formation activity), we measure the neighbour density distributions around isolated galaxies (Fig.~\ref{fig:orbits_schematic}; described in full detail in \S\ref{sec:methods}). We call these isolated galaxies our \textit{isolated host} sample. In \S\ref{sec:obs_spec}, we describe the spectropsic data from the Sloan Digital Sky Survey \citep[SDSS;][]{SDSS_DR16} used to identify these isolated hosts, and in \S\ref{sec:obs_phot}, we describe the photometric data from the SDSS used to measure the density distribution of nearby neighbours. To demonstrate that our technique can test whether we observe positive correlations between star formation and dark matter accretion, we also use simulated data from the \UM{} model \cite{Behroozi19}; this data is described in \S\ref{sec:UM}.

\subsection{Spectroscopic Data (Observed Isolated Hosts)}
\label{sec:obs_spec}

To identify isolated hosts from SDSS, we use data from the DR16 spectroscopic catalogues \citep{SDSS_DR16}, which are $> 90\%$ complete for galaxies brighter than $r = 17.77$. Following the procedure in \cite{Behroozi15},we use median stellar masses and specific star formation rates from the MPA-JHU value-added catalogue \citep{Kauffmann03, Brinchmann04}. These values were calculated assuming a \cite{Kroupa02} initial mass function (IMF), and we convert them to a \cite{Chabrier03} IMF. For galaxies with fibre collisions, we supplemented the catalogue with data from the NYU Value-Added Galaxy Catalog \citep[NYU-VAGC;][]{Blanton} for galaxies with $\log_{10}(M_*/M_\odot) > 9.5$ to improve our isolated hosts selection. However, our results do not change without the addition of the galaxies from the NYU-VAGC.

Our spectroscopic catalogue covers an on-sky area of 8,427.7 deg$^2$ and includes 697,477 galaxy targets with nonzero stellar masses. To apply our isolation criterion, we exclude galaxies that are within 2 Mpc of a survey boundary or a region of significant incompleteness. We also exclude galaxies with $z < 0.01$ to avoid Hubble flow corrections \citep[e.g.,][]{Baldry12}. Our resulting catalogue has 547,271 galaxies over 6401.1 deg$^2$ of sky.

Since the SDSS is magnitude-limited, we perform cuts to convert our spectroscopic catalogue to a stellar mass-complete sample. Following \cite{Behroozi15}, over 95\% of galaxies with a given stellar mass $M_*$ at redshift $z$ satisfy
\begin{equation}
    r < -0.25 - 1.9 \log_{10} \left( \frac{M_*}{M_\odot} \right) + 5 \log_{10} \left( \frac{D_L(z)}{10\mathrm{pc}} \right)
    \label{eq:SM_complete_cut}
\end{equation}
in the SDSS, where $r$ is the galaxy's $r$-band apparent magnitude and $D_{L}$ is the luminosity distance given our cosmology. Given SDSS's spectroscopic survey limits, we exclude galaxies for which $r > 17.77$ according to Eq.~\ref{eq:SM_complete_cut}, since it would be otherwise impossible to apply our isolation criteria.

Finally, since a purely volume-limited catalogue would unacceptably reduce the size of our isolated galaxy sample, we weight neighbour density distributions by the inverse of the observable volumes for each isolated galaxy, obtained by inverting Eq.~\ref{eq:SM_complete_cut} with $r = 17.77$. We discuss our weighting of our observational data in more detail in  \S\ref{sec:bkgd_weight}.

\subsubsection{Sample Statistics}
\label{sec:SDSS_sample_stats}

We identify 25,625 isolated galaxies\footnote{The isolated galaxy catalog is available at \url{https://github.com/caodonnell/DM_accretion}} from SDSS within a redshift range of $0.01 < z < 0.123$. To measure the uncertainty in neighbour density distributions, we used 112 jackknife samples. For each jackknife sample, a $\sim 10^\circ \times 10^\circ$ ($\sim 37.5 \times 37.5$ Mpc/$h$ at the median host redshift $z=0.079$) region was removed from the sky footprint for the analysis, resulting in an average of $\sim$25,400 isolated hosts per jackknife sample.

\subsubsection{Star-forming \& Quiescent Bins}
\label{sec:sf_vs_q}

\begin{figure}
    \centering
    \includegraphics[width=\columnwidth]{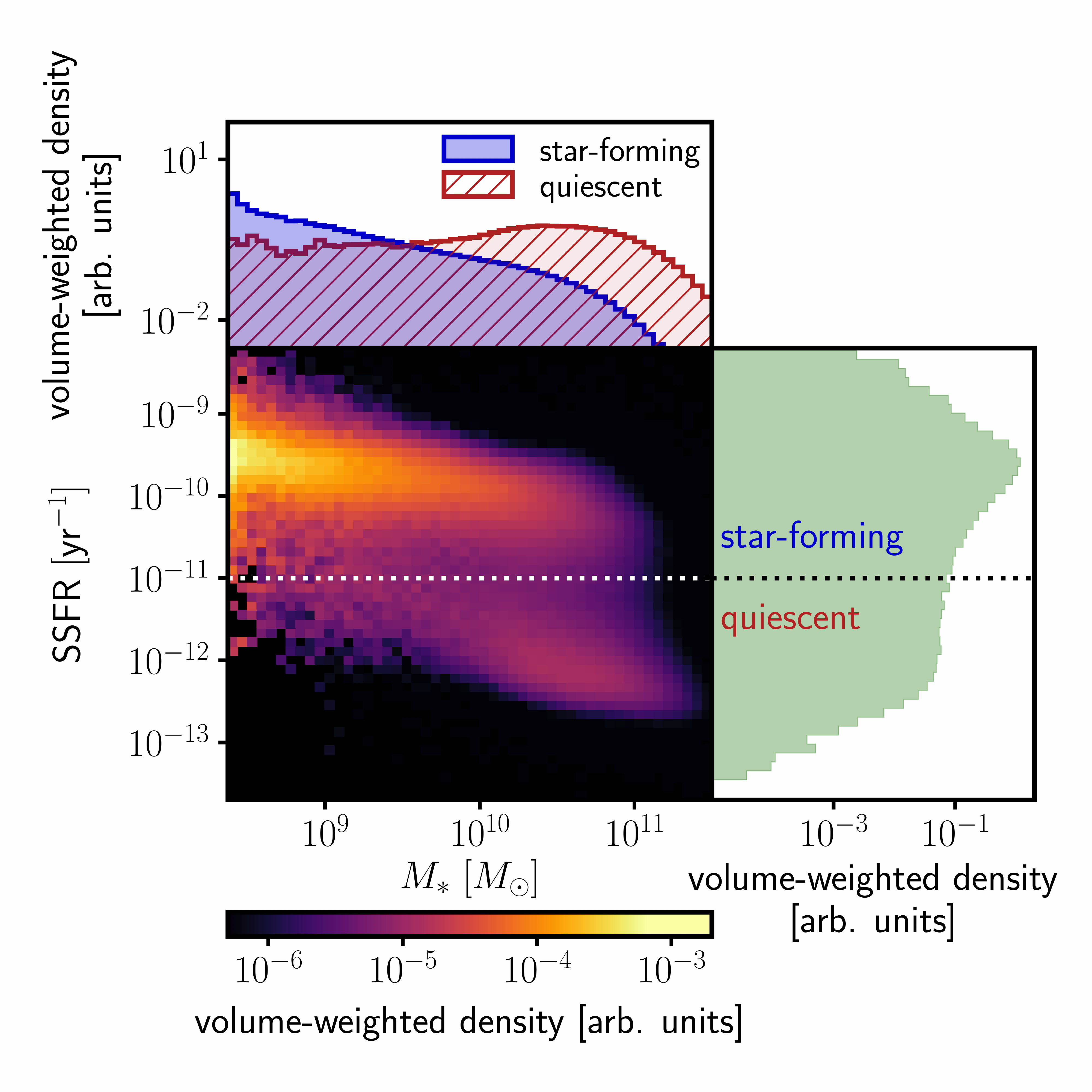}\\[-0.5ex]
    \vspace{-\baselineskip}
    \caption{We identified star-forming versus quiescent galaxies based on their specific star formation rates (SSFRs); following \protect\cite{Wetzel12}, star-forming hosts were those with $\mathrm{SSFR} > 10^{-11} \, \mathrm{yr}^{-1}$. The central plot shows the volume-weighted density distribution of galaxies in the DR16 spectroscopic catalogue (\S\ref{sec:obs_spec}). The histogram above shows the distribution of stellar masses of star-forming versus quiescent hosts, and the histogram to the right shows the overall distribution of specific star formation rates with the split at $10^{-11} \, \mathrm{yr}^{-1}$ indicated.}
    \label{fig:SM_SSFR}
\end{figure}

\begin{figure}
    \centering
    \includegraphics[width=\columnwidth]{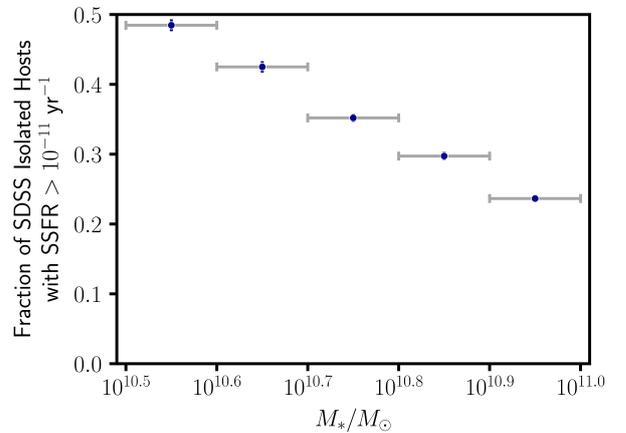}\\[-1ex]
    \vspace{-\baselineskip}
    \caption{The fraction of star-forming isolated hosts (i.e., those with $\mathrm{SSFR} > 10^{-11} \, \mathrm{yr}^{-1}$) ranges from $48\%$ to $24\%$ across the isolated host stellar mass range. Each fraction indicates the fraction of star-forming hosts within a 0.1 dex bin (e.g., from $10.5 < \log_{10}(M_*/M_\odot) < 10.6$). The vertical blue error bars indicate the scatter in the fraction across the jackknife samples, and the grey bars indicate the stellar mass bin width.}
    \label{fig:sf_fraction}
\end{figure}

To constrain the correlation between dark matter accretion rates and star formation rates, we split our sample of isolated hosts from the SDSS into star-forming and quiescent bins based on their specific star formation rates (SSFRs). Following \cite{Wetzel12}, we separated the bins at $\mathrm{SSFR} = 10^{-11} \, \mathrm{yr}^{-1}$ (Fig.~\ref{fig:SM_SSFR}). Across the entire isolated host mass range from $10.5 < \log_{10}(M_*/M_\odot) < 11.0$,  the fraction of isolated hosts that were star-forming ranged from $48\%$ to $24\%$ (Fig.~\ref{fig:sf_fraction}), and we apply these fractions to the simulation data as described in \S \ref{sec:UM}. The redshift distributions of the star-forming isolated hosts and quiescent isolated hosts are similar, and both have an average redshift $z=0.074$.

\subsection{Photometric Data (Observed Nearby Neighbours)}
\label{sec:obs_phot}

\begin{figure*}
    \centering
    \includegraphics[width=\textwidth]{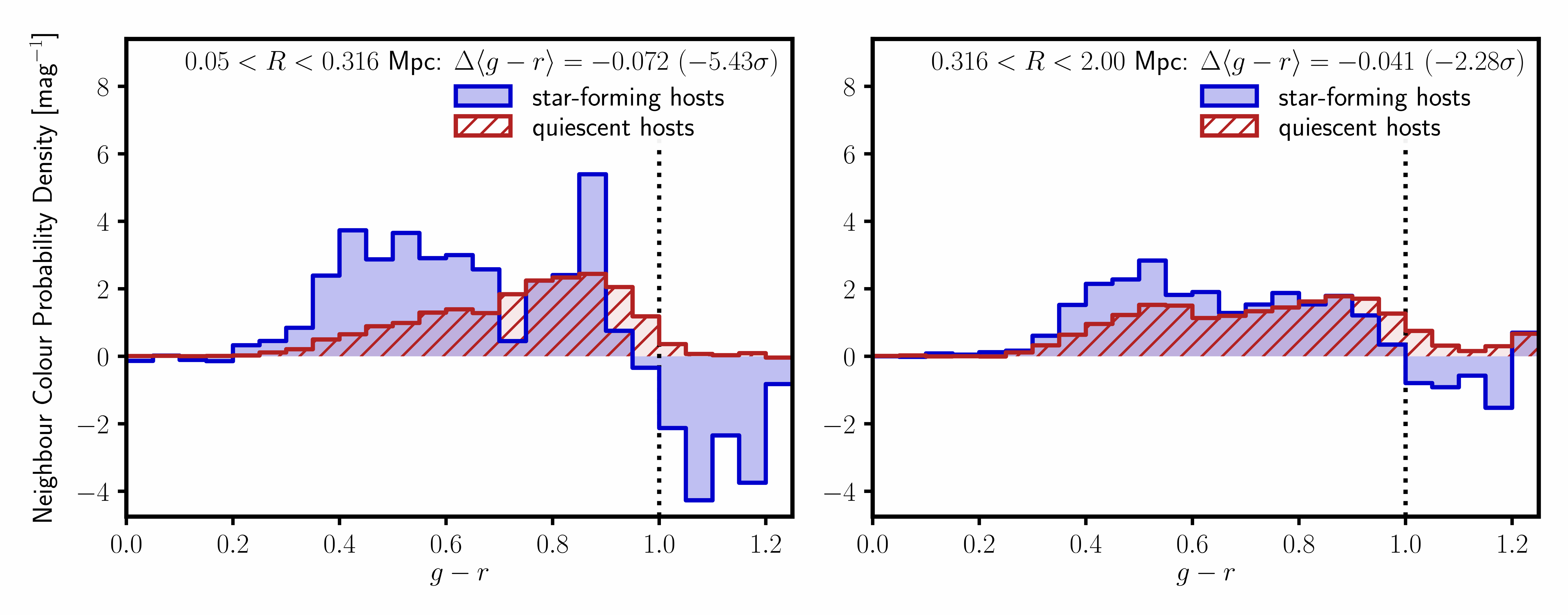}\\[-1ex]
    \vspace{-\baselineskip}
    \caption{Because galaxies with redder colours tend to be at higher redshifts, we reduce noise by applying a colour cut of $g-r < 1.0$ to photometric galaxies; the upper limit is indicated by the black dotted vertical line. These plots include neighbours with $\log_{10}(M_*/M_\odot) > 9.36$, which corresponds to the stellar mass limit at the maximum isolated host redshift ($z=0.123$) given SDSS photometric limits. The projected distance ranges of the two panels are set to match those used in our analysis of the shape of the neighbour density distribution (\S\ref{sec:shape_ratio}).  Close to star-forming host galaxies, neighbours have bluer $g-r$ colours than neighbours around quiescent galaxies, but further from the isolated hosts, the colour differences for neighbours around star-forming and quiescent host galaxies are less significant.}
    \label{fig:Mstellar_gr_hist}
\end{figure*}

To count neighbours around each isolated galaxy, we use sky position and $g-r$ colours from the DR16 photometric catalogue \citep{SDSS_DR16}. We only use sources with $r<21.5$ so that galaxies with $g-r$ colours as red as 1.25 are still above the SDSS $g$ band sensitivity (90\% sensitivity: $g>22.78$). Additionally, we only use sources with a type field of `GALAXY' to exclude likely stars. The full catalogue contains 73,109,495 galaxies with $r < 21.5$ over an on-sky area of 18,509.0 deg$^2$. 

Previous studies have binned neighbours according to their luminosity \citep[e.g.,][]{More16, Baxter17}. However, the  satellites of star-forming host galaxies are expected to be brighter and bluer because they are more often star-forming than the satellites around quiescent hosts \citep[e.g.,][]{Weinmann06, Berti17}. Fig.~\ref{fig:Mstellar_gr_hist} confirms this bias in our sample: close to star-forming hosts, neighbours tend to be bluer than neighbours close to quiescent hosts ($\Delta \langle g-r \rangle = -0.072 = -5.43\sigma$ via bootstrapping), which is consistent with \cite{Weinmann06}. This trend may continue further from the isolated hosts ($\Delta \langle g-r \rangle = -0.041$), but the significance is lower ($-2.28 \sigma$). Our analysis relies on the shape of the neighbour density distribution (\S\ref{sec:shape_ratio}), but binning neighbours by luminosity may affect the shape of these distributions. For example, a luminosity-based binning scheme may be biased against the fainter satellites around quiescent galaxies, resulting in a shallower neighbour density distribution.

For our analysis, we instead bin neighbours by their stellar masses, which are expected to be more consistent throughout satellite galaxy orbits. We assume that all nearby neighbours are at the same redshifts as their hosts for calculating their stellar masses, using background subtraction to remove background and foreground contamination \citep[see e.g.,][]{Lan16}.

\begin{figure}
    \centering
    \includegraphics[width=\columnwidth]{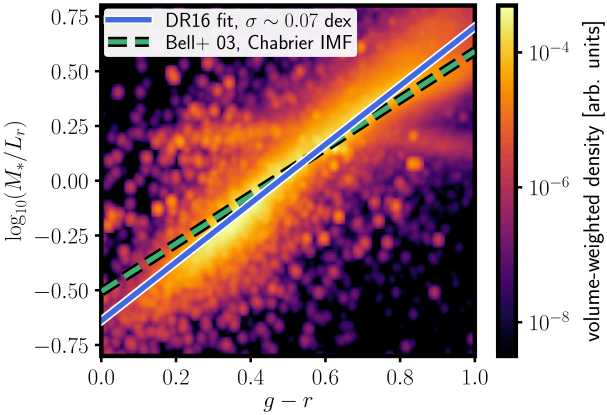}\\[-1ex]
    \vspace{-\baselineskip}
    \caption{We fit a relation between galaxies' $g-r$ colours and their $M_*/L_r$ ratios following \protect\cite{Bell03}. This allows us to bin nearby photometric neighbours by stellar mass instead of luminosity, as stellar mass should be more robust throughout a satellite galaxy's orbit. Our fit had a scatter of $\sigma \sim 0.07$ dex. The \protect\cite{Bell03} line includes a normalisation factor to convert its IMF to be consistent with our SDSS DR16 data (-0.2 dex following \protect\cite{Salim07}). The difference between the two fits are due to different assumptions, including dereddened versus $k$-corrected colours and the redshift ranges of included galaxies (\S\ref{sec:obs_phot}).  To reduce visual noise, the density plot shown above is smoothed with a Gaussian kernel of width 0.07 dex in mass-to-light ratio and 0.04 mag in colour.}
    \label{fig:Mstellar_gr_fit}
\end{figure}

Following \cite{Bell03}, we used linear regression to fit a relation for mass-to-light ratios as a function of the $g-r$ colours using galaxies from the SDSS DR16 spectroscopic catalogue \citep{SDSS_DR16}. We selected spectroscopic galaxies that match the properties of our expected nearby neighbours, i.e., $0.01 < z < 0.125$ and $8.0 < \log_{10}(M_*/M_\odot) < 11.0$, and weighted by stellar mass completeness (Eq. \ref{eq:w_vol}). We find a best fit of
\begin{equation}
    \log_{10}(M_*/L_r) = 1.341 \, (g-r) - 0.639 \, ,
    \label{eq:MsLr_gr}
\end{equation}
with a scatter $\sigma \sim 0.07$ dex (Fig.~\ref{fig:Mstellar_gr_fit}). We note that our fit differs from the fit found in \cite{Bell03}, due to the different assumptions used. First, \cite{Bell03} used stellar masses derived from a `diet' \cite{Salpeter55} IMF, whereas our stellar masses come from \cite{Brinchmann04} converted to a \cite{Chabrier03} IMF. Additionally, we restricted our fit to galaxies with stellar masses that correspond to stellar mass bins used in our analysis. In Fig.~\ref{fig:Mstellar_gr_fit} below, we convert the fit from \cite{Bell03} to a Chabrier IMF by including a normalisation factor of -0.2 dex following \cite{Salim07}. Second, \cite{Bell03} used galaxies from the SDSS Early Data Release \citep{SDSS_EDR}, which included galaxies in the redshift range $0.0< z < 0.5$. For our fit, we only consider galaxies in a smaller redshift range from $0<z<0.125$. Finally, because of our restricted redshift range, we perform our fit with dereddened colours from the SDSS photometric catalogues and did not apply $k$-corrections, but \cite{Bell03} used $k$-corrected colours. In Appendix \ref{sec:appendix}, we consider neighbour density distributions resulting from binning neighbours by luminosity (\S\ref{app:Ms_Mr_bins}) and using the stellar mass fit from \cite{Bell03} (\S\ref{app:Ms_gr_fit_Bell}), and we find no differences in our conclusions.

When applying the fit in Eq. \ref{eq:MsLr_gr} to our photometric catalogue, we required that galaxies have $0.0 < g-r < 1.0$ to ensure reliable photometry and to exclude galaxies at higher redshifts. We determined this cut from the colour distributions of nearby neighbours around isolated hosts (Fig.~\ref{fig:Mstellar_gr_hist}), which shows that galaxies with $g-r > 1.0$ are consistent with background noise. These redder galaxies are expected to be at higher redshifts; we have tested that repeating our analysis with a redder colour cut of $g-r > 1.25$ yields the same results. After these cuts, our photometric sample contained 35,457,243 galaxies.

\subsection{Simulation Data}
\label{sec:UM}

\subsubsection{Overview}

To demonstrate our analysis technique, we use simulation data. The haloes are from the \textit{Bolshoi-Planck} dark matter simulation \citep{Klypin-BP, RodriguezPuebla-BP} with galaxy stellar masses from the \UM{} empirical model \citep{Behroozi19}. The \textit{Bolshoi-Planck} simulation had a co-moving volume of (250 Mpc$/h$)$^3$ with 2048$^3$ particles ($\sim 8 \times 10^9$) with high mass resolution ($1.6 \times 10^8 h^{-1}  \textrm{M}_\odot$). They adopted a flat $\Lambda$CDM cosmology ($h = 0.678$, $\Omega_m = 0.307$, $\sigma_8 = 0.823$, $n_s = 0.96$) that is compatible with  \textit{Planck} 2015 and 2018 results \citep{Planck15, Planck18}; we use this same cosmology in this paper. Halo finding and merger tree construction were done using \textsc{Rockstar} \citep{Behroozi_rockstar} and \textsc{Consistent Trees} \citep{Behroozi_consistenttrees} codes, respectively.  Halo masses ($M_\mathrm{vir}$) were defined using the virial spherical overdensity criterion ($\rho_\mathrm{vir}$) of \cite{Bryan98}.

In this paper, we use halo accretion rates from \textit{Bolshoi-Planck} over the past dynamical time $t_{\mathrm{dyn}} = 1/\sqrt{G\rho_{\mathrm{vir}}}$, normalised by halo virial masses, i.e.,
\begin{equation}
\Gamma = \frac{\Delta \log (M_\mathrm{vir})}{\Delta \log (a)},
%\frac{M_{\mathrm{vir}}(t_{\mathrm{now}}) - M_{\mathrm{vir}}(t_\mathrm{now} - t_\mathrm{{dyn}})}{M_{\mathrm{vir}}(t_{\mathrm{now}})} \, . 
\label{eq:spec_acc_rate}
\end{equation}
following \cite{Diemer14}. These are \textit{specific} halo mass accretion rates, and their distribution depends only weakly on halo mass \citep{Behroozi15b}. We also tested the \UM{} data using specific halo accretion rates over the past $2t_\mathrm{dyn}$, which did not change our conclusions and is presented in Appendix \ref{app:2tdyn}. Furthermore, to ensure our results are not impacted by stacking data from correlated snapshots, we also did our analysis using haloes from a single snapshot of the Small MultiDark Planck simulation \citep[SMDPL;][]{Klypin16} with $a=0.956$ in Results \S~\ref{sec:results_UMoffset}. Compared with the \textit{Bolshoi-Planck} simulation, the SMDPL has a larger co-moving volume of (400 Mpc$/h$)$^3$, which gives us greater precision for an analysis that relies on a single snapshot.

The \UM{} is an empirical model that uses a Markov Chain Monte Carlo (MCMC) algorithm to model the relationships between galaxy properties and dark matter halo properties \citep{Behroozi19}. This model uses halo properties and assembly histories from the \textit{Bolshoi-Planck} simulation, and it self-consistently constrains individual galaxies' properties to match observed stellar mass functions ($z \sim 0- 4$), cosmic star formation rates ($z\sim 0 -10$), specific star formtion rates ($z \sim 0-8$), UV luminosity functions ($z \sim 4 -10$), quenched fractions ($z\sim 0-4$), auto- and cross-correlation functions ($z \sim 0-0.5$), and median UV-stellar mass relations ($z \sim 4-10$); full references are available in appendix C of \cite{Behroozi19}. Of note, the \UM{} model allows for orphans, i.e., satellites are allowed to persist after being destroyed in the dark matter simulation. Without orphans, the predicted spatial correlation of galaxies is much lower than observed. More details are in appendix C of \cite{Behroozi19} and \S{}2.2.2. of \cite{Allen19}.

In the \UM{}, star formation rates are parameterised as a function of halo mass, halo accretion rates, and redshift. Stellar masses at $z=0$ are constrained to match \cite{Moustakas13} with corrections for extended galaxy profiles as described in \cite{Bernardi13}. We note that these masses differ from stellar masses in SDSS due to the treatment of galaxy light profiles, and we describe this systematic in more detail in \S\ref{sec:mass_fcn}. We adjust stellar masses from the \UM{} to match the calibration used for SDSS stellar masses for consistency (\S\ref{sec:mass_fcn}).  Observed stellar masses from the \UM{} incorporate both (1) systematic offsets between true and observed stellar masses as well as (2) random scatter in observed stellar masses. The resulting observables from the \UM{} data used in this paper include galaxy positions, velocities, and stellar masses. We do not use the star formation rates generated by the \UM{} because we instead use halo mass accretion rates as described below in \S\ref{sec:UM_accrates}.

\subsubsection{Sample Statistics}
\label{sec:UM_sample_stats}

We combined \UM{} simulation data from 14 snapshots with $a = 0.904$ to $a = 1.002$. Each snapshot had an average of 31,026 isolated hosts\footnote{The isolated host catalog is available at \url{https://github.com/caodonnell/DM_accretion}}. $96.5\%$ of the isolated haloes were not satellites of larger haloes. 

We created 25 jackknife samples by leaving out a 50 $\times$ 50 Mpc/$h$ region along the $x$ and $y$ axes, taking the $z$ axis as our line-of-sight. For each jackknife sample, we stacked the neighbor density distributions from the \UM{} data by averaging across the 14 snapshots, leaving out the same region from each snapshot. Each snapshot contained an average of $\sim$27,000 isolated hosts per sample. We note that the uncertainties for the SDSS and \UM{} neighbour density distributions are different because the background (noise) from the SDSS photometric data includes galaxies out to $z \sim 0.2$ (over 570 Mpc/$h$) whereas the \UM{} simulation box is only 250 Mpc/$h$ per side.

\subsubsection{Correlating Star Formation Activity with Halo Mass Accretion Rates}
\label{sec:UM_accrates}

The \UM{} generates star formation rates based on an assumption about the correlation between star formation rates and host halo accretion rates. However, in this paper, we want to measure this correlation, and so we discard the star formation rates from the \UM{}. Instead, we categorise galaxies as being star-forming or quiescent based only on their host halo dark matter specific accretion  (Eq.~\ref{eq:spec_acc_rate}).

To predict observable effects from correlations between accretion rates and star formation, we constructed analogues of the star-forming and quiescent SDSS host galaxies from the \UM{} data. As described in \S\ref{sec:sf_vs_q}, the fraction of star-forming isolated galaxies in the SDSS ranges from 48 to 24\% across the host stellar mass range. Within each equivalent isolated host stellar mass bin from \UM{} data, we split the hosts into a high-accreting host subsample and a low-accreting host subsample. For positive correlation strengths, the fraction of \UM{} hosts in the high-accreting host subsample is set to match the fraction of star-forming hosts in the corresponding SDSS host stellar mass bin, whereas for negative correlation strengths, the fraction of \UM{} hosts in the low-accreting host subsample is set to match the fraction of star-forming hosts in the corresponding SDSS host stellar mass bin. 

When predicting the neighbour density distributions for different correlations ($\rho$), we select hosts such that a fraction $\rho$ are chosen from the corresponding host subsample (high-accreting or low-accreting) and the remaining fraction $1-\rho$ are chosen randomly from all isolated hosts. For example, each snapshot from the \UM{} has $\sim$27,000 isolated host galaxies (\S\ref{sec:UM_sample_stats}) of Milky Way-mass. Of those hosts, $\sim$18,750 have $10.7 < \log_{10}(M_*/M_\odot) < 10.8$. For the isolated hosts from the SDSS within that mass range, 35\% were star-forming (Fig.~\ref{fig:sf_fraction}). Thus, for positive correlations, we first split the $\sim$5,500 hosts into (1) a high-accreting subsample with the 35\% of host haloes with the highest accretion rates and (2) a low-accreting subsample with the remaining 65\% of host haloes with lower accretion rates. For an example correlation rate of $\rho = 0.50$, the star-forming hosts in the \UM{} with $10.7 < \log_{10}(M_*/M_\odot) < 10.8$ consisted of $\sim$1,925 hosts (35\% of the total number of hosts in the mass range) where half ($\rho = 0.5$) were randomly selected with replacement from the high-accreting host subsample and the other half ($1-\rho = 0.5$) were randomly selected with replacement from all hosts in the mass range. Finally, the quiescent hosts from the \UM{} with $10.7 < \log_{10}(M_*/M_\odot) < 10.8$ consisted of $\sim$3,575 hosts (65\% of the total number of hosts in the mass range), and for $\rho=0.50$, half were randomly selected with replacement from the low-accreting host subsample and the other half were randomly selected with replacement from all the hosts in the mass range.

\begin{figure}
    \centering
    \includegraphics[width=\columnwidth]{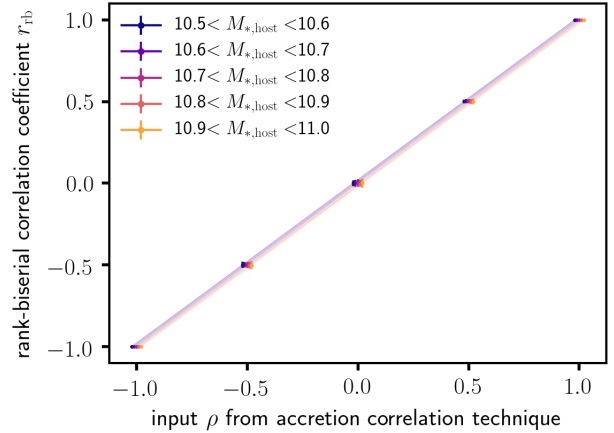}\\[-2ex]
    \caption{Our technique for simulating different correlation strengths between star formation and dark matter accretion in the \UM{} effectively recovers expected correlation coefficients. This plot compares the input correlation strength $\rho$ following the technique described in \S\ref{sec:UM_accrates} versus the computed rank-biserial correlation coefficients. An arbitrary offset in the $x$ direction has been added to separate the points, and each line represents a 0.1-dex bin in isolated host stellar mass.}
    \label{fig:accrate_correlation}
\end{figure}

To confirm that this sampling technique effectively measures correlation strength between (1) the binary category of star-forming versus quiescent hosts and (2) the hosts' specific halo accretion rates, we computed the rank-biserial correlation coefficent $r_\mathrm{rb}$ for our \UM{} samples. As depicted in Fig.~\ref{fig:accrate_correlation}, our technique successfully recovers the expected correlation strengths.

\section{Methods}
\label{sec:methods}

In \S\ref{sec:methods_overview}, we discuss how we identify isolated host galaxies and measure their neighbour density distributions. Next, in \S\ref{sec:shape_ratio}, we describe how we measure the correlation between star formation and accretion rates. Finally, in \S\ref{sec:systematics}, we describe systematics addressed in our analysis.

\subsection{Measuring Neighbour Density Distributions around Isolated Host Galaxies}
\label{sec:methods_overview}

Our method measures the average density distribution of neighbouring galaxies around galaxies with stellar masses $10.5 < \log_{10}(M_*/M_\odot) < 11.0$, corresponding to haloes with masses $\sim{}10^{12} - 10^{13}M_\odot$ \citep{Behroozi19}. We specifically target isolated galaxies to eliminate contamination from satellites of larger nearby haloes. We consider a galaxy `isolated' if no larger galaxy is found within 2 Mpc projected (on-sky) physical distance and 1000 km/s velocity distance.  In the \UM{}, $>$96\% of galaxies passing this cut are central galaxies (i.e., not satellites of a larger halo). We term these galaxies our \textit{isolated host} sample. 

\begin{figure}
    \centering
    \includegraphics[width=\columnwidth]{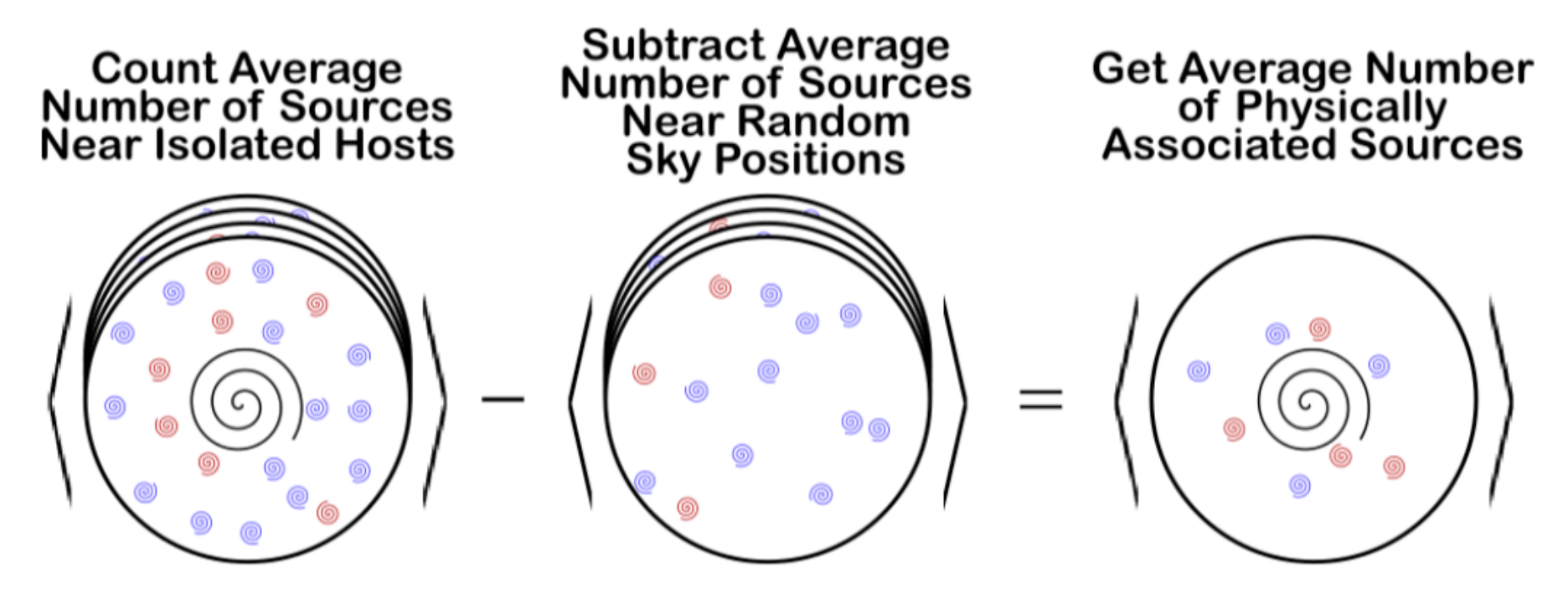}
    \vspace{-\baselineskip}
    \caption{Schematic of the background subtraction technique used in this work. For each host galaxy, 100 random pointings are selected following the same iso lation criteria. By subtracting the average neighbour density distribution around the randoms from the average distribution around the host galaxies, we recover the average density distribution of physically associated sources.}
    \label{fig:method_cartoon}
\end{figure}

To subtract foreground/background sources, we select 100 random pointings for each host galaxy following the same isolation criteria within the same sky footprint. We then count the number of neighbouring galaxies in annuli around each host galaxy and random pointing. As depicted in Fig.~\ref{fig:method_cartoon}, by subtracting the neighbour distribution around the random pointings from the distribution around host galaxies, we measure the average number density distribution of physically associated neighbours. This same technique is used in \cite{More16} and \cite{Baxter17}. We test several mass thresholds for nearby neighbours to verify that the mass threshold used does not impact the results.

The outermost annulus radius for counting neighbouring galaxies is 2 Mpc, consistent with our isolation criteria. The innermost annulus radius is 50 kpc to avoid possible influence of the host galaxy on source detection in the SDSS data. We checked this limit by conducting our analysis with both SDSS DR16 \citep{SDSS_DR16} and DR7 \citep{SDSS_DR7} photometric catalogues. Close to isolated hosts, we could have systematic biases between the number density of galaxies around hosts versus the number density around random pointings. Between DR7 and later releases, the background subtraction algorithm in SDSS was improved \citep{Blanton11}. Our results are consistent between DR16 and DR7, and we would expect that the improvement in SDSS background subtraction should be larger than any remaining systematics. 

We then stack the background-subtracted neighbour density distributions of the isolated host galaxies in our sample. To account for completeness of the SDSS sample, we weight each distribution by the inverse of the co-moving volume out to which the host could be observed based on its stellar mass and SDSS spectroscopic survey limits; \S{}\ref{sec:bkgd_weight} below provides more details on the weighting applied to the SDSS sample. To estimate the uncertainty in the density distributions, we perform a jackknife statistical analysis on observational and theoretical samples. \S\ref{sec:SDSS_sample_stats} and \S\ref{sec:UM_sample_stats} above discussed sample statistics for the observation and simulation data, respectively.

\subsection{Constraining Accretion Rate Correlations Using Neighbour Distributions}
\label{sec:shape_ratio}

\begin{figure*}
    \centering
    \includegraphics[width=\textwidth]{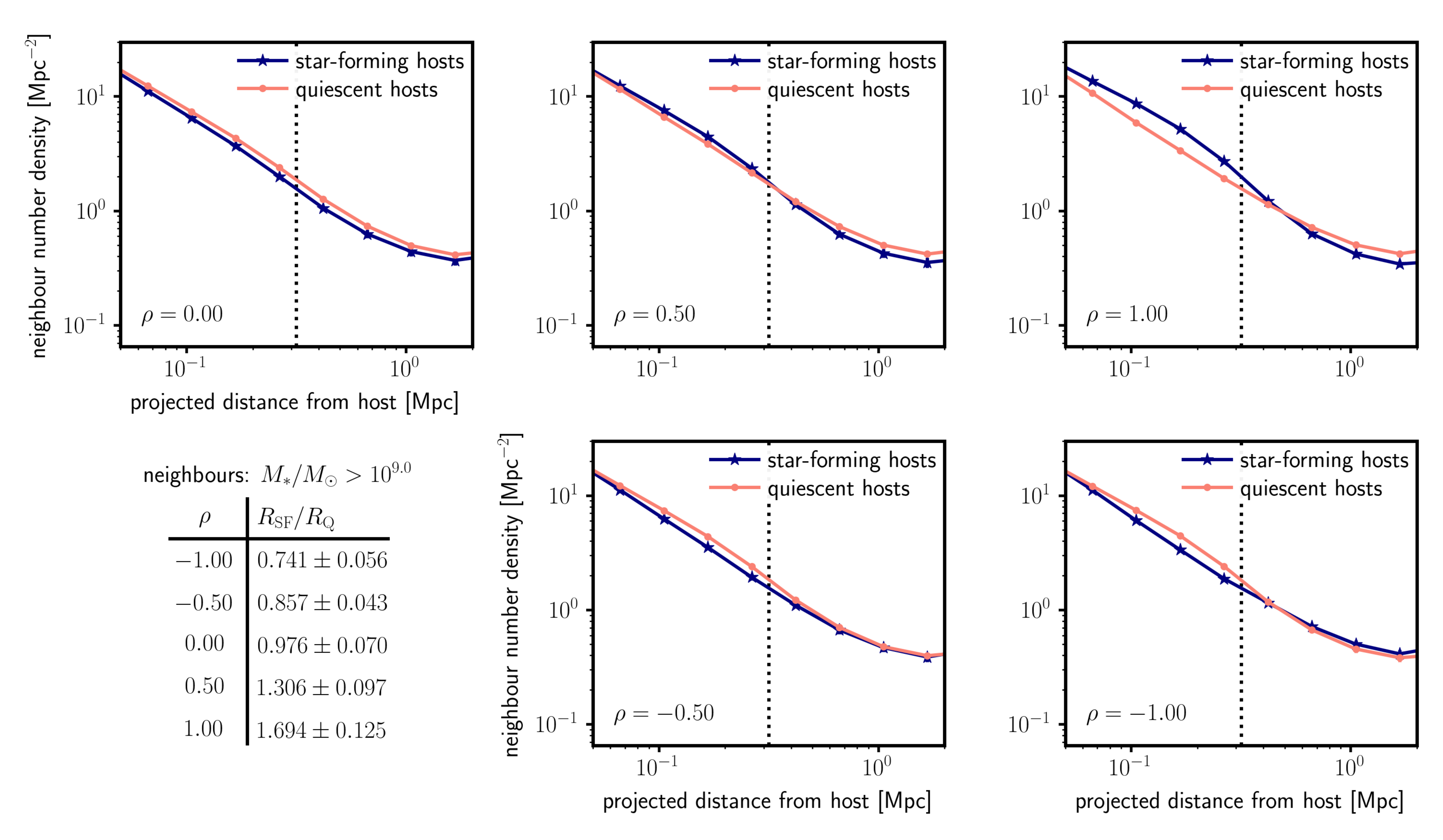}%\\[-2ex]
    \vspace{-\baselineskip}
    \caption{Stronger positive correlations between accretion rates and star formation rates lead to steeper neighbour galaxy profiles around star-forming galaxies in the \UM{} simulations. We measure the shapes of the neighbour density distributions using a shape parameter to compare the inner ($0.05 < r < 0.316$ Mpc) and outer neighbour counts ($0.316 < r < 2.0$ Mpc; Eq.~\ref{eq:shape_ratio}). The \textcolor{blue}{blue lines} represent the analogues to the star-forming galaxies from the SDSS, and the \textcolor{red}{red lines} represent the analogues to the quiescent SDSS galaxies. The error bars represent the scatter across jackknife samples, and the \textbf{dashed vertical lines} represent $r_\mathrm{split}=0.316$ Mpc used in the shape parameter calculations. In these plots, the neighbour number density includes neighbours with $\log_{10}(M_*/M_\odot) > 9.0$. The top three panels depict different correlation strengths between dark matter accretion rates and SSFR (0\%, 50\%, and 100\% from left to right), and the bottom two panels depict negative correlation strengths (-50\% and -100\% from left to right). The inset table indicates the shape ratio (\S\ref{sec:shape_ratio}) for each panel, which compares the shape parameters (Eq.~\ref{eq:shape_ratio}) for the distributions. In the $\rho=0.0$ case (no correlation), the offset in the neighbour density distributions between star-forming and quiescent hosts is due to the quiescent sample having larger host halo masses.}
    \label{fig:accret_rate}
\end{figure*}

As a halo accretes more material, its gravitational potential well deepens, and this change will impact the orbits of satellite galaxies (Fig.~\ref{fig:orbits_schematic}). As described in \cite{Diemer13} and \cite{WetzelNagai15}, because dark matter is dissipationless, it will be deposited in a shell-like manner at $\sim R_{200\mathrm{m}}$. As a result, \cite{Diemer13} found little ($\sim$10\%) growth in halo mass at smaller radii between $z=1$ to $z=0$. Similarly, when comparing the neighbour density distributions around isolated haloes assuming different correlation strengths (Fig.~\ref{fig:accret_rate}), we find a steepening in the profile at a few hundred kpc, which corresponds to $R_{200\mathrm{m}}$ for our halo masses at $z=0$. At smaller distances, highly-accreting hosts pull satellites inwards and can tidally disrupt these galaxies. At large distances (i.e., beyond the virial radius), the distributions are more similar.

\begin{figure}
    \centering
    \includegraphics[width=\columnwidth]{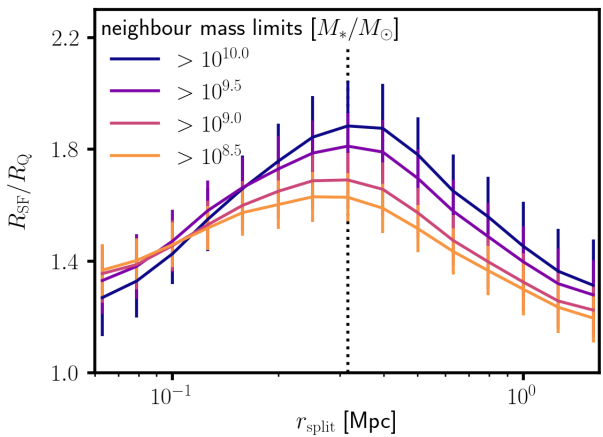}\\[-0.5ex]
    \vspace{-\baselineskip}
    \caption{The shape ratio is most sensitive to accretion rate differences when the inner versus outer regions for the shape parameters (Eq.~\ref{eq:shape_ratio}) are split around 0.316 Mpc (indicated by the dotted line), and this value does not depend on the choice of neighbour mass selection. This plot assumes 100\% correlation between accretion rates and star formation rates; lower correlations give identical results for the optimal choice of $r_{\mathrm{split}}$.}
    \label{fig:shape_ratio}
\end{figure}

To assess the shape of the density distribution of neighbouring haloes, we define a \textit{shape parameter} which compares the number of neighbours close to the host versus further from the host, specifically
\begin{equation}
    R = \frac{N \in (0.05 \textrm{ Mpc} < r < r_{\mathrm{split}})}{N \in (r_{\mathrm{split}} < r < 2.0  \textrm{ Mpc})} \, .
    \label{eq:shape_ratio}
\end{equation}
where the innermost radial distance (0.05 Mpc) is set to conservatively exclude incompleteness from source blending in the SDSS data, and the outer limit (2.0 Mpc) is matched to our isolation criteria (\S\ref{sec:methods_overview}). We find that $r_{\mathrm{split}} \equiv 0.316$  Mpc maximises the differences between the neighbour density distributions around high-accreting versus low-accreting hosts (Fig.~\ref{fig:shape_ratio}). We quantify these differences with a \textit{shape ratio} $R_{\mathrm{SF}}/R_{\mathrm{Q}}$, which is the ratio of the shape parameters for the neighbour density distributions around star-forming hosts ($R_{\mathrm{SF}}$) versus quiescent hosts ($R_{\mathrm{Q}}$). The choice of neighbour mass limit does not affect the choice of $r_{\mathrm{split}}$. Additionally, we do not find that metric is significantly impacted by the \UM{} orphan model (Appendix \ref{app:orphan}). We use the shape ratio as the metric throughout the rest of our analysis. 

To demonstrate this approach, we apply our technique to the neighbour density distributions around isolated hosts in the \UM{}. As shown in Fig.~\ref{fig:accret_rate}, $R_\mathrm{SF}/R_\mathrm{Q} > 1$ implies positive correlations ($\rho > 0$) between dark matter accretion and star formation rates, whereas $R_\mathrm{SF}/R_\mathrm{Q} < 1$ implies negative correlations ($\rho < 0$).

\subsection{Systematics and Selection Effects}
\label{sec:systematics}
Below, we describe systematics and selection effects that differ between the \UM{} and SDSS data. We address overall offsets in stellar mass definitions  (\S\ref{sec:mass_fcn}), offsets in stellar masses between star-forming and quiescent galaxies (\S\ref{sec:density_offset}), and weighting for completeness as well as maximizing signal-to-noise (\S\ref{sec:bkgd_weight}).

\subsubsection{Stellar Mass Functions}
\label{sec:mass_fcn}

\begin{figure}
    \centering
    \includegraphics[width=\columnwidth]{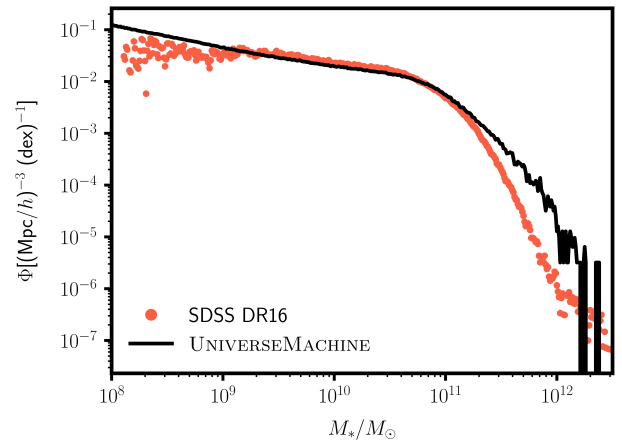}\\[-0.5ex]
    \vspace{-\baselineskip} 
    \caption{The differences between the \UM{} and SDSS MPA-JHU mass functions are largely due to different ways of determining galaxy luminosities and masses (i.e., \citealt{Bernardi13} vs.\ \citealt{Kauffmann03}). To account for the differences, we use mass cutoffs in the \UM{} catalogues such that the cumulative number densities of more massive objects match that for the SDSS (Table~\ref{tab:MF}).}
    \label{fig:MF}
\end{figure}

\begin{table}
\begin{center}
\begin{tabular}{c|c|c|c}
    &SDSS  & \UM{} & $\Phi(>M_*)$ \\
    &$\log_{10}(M_*/M_\odot)$ & $\log_{10}(M_*/M_\odot)$ & (Mpc/$h$)$^{-3}$ (dex)$^{-1}$ \\
    \hline 
    \multirow{2}{*}{\rotatebox{90}{Hosts}} 
    & 10.50 & 10.50 & 0.64016  \\
& 11.00 & 11.08 & 0.09464  \\
\hline \hline
\multirow{4}{*}{\rotatebox{90}{Neighbours}}
&8.50 & 8.62 & 6.62222  \\
&9.00 & 8.93 & 4.85279  \\
&9.50 & 9.38 & 3.05361  \\
&10.00 & 9.93 & 1.62929 \\
\hline
\end{tabular}
\caption{As discussed in \S \ref{sec:mass_fcn}, stellar mass definitions differ between the SDSS MPA-JHU and \UM{}.  This table summarises analogous stellar masses between the SDSS and the \UM{} (first two columns, respectively) based on matching the cumulative number density of more massive galaxies (third column). The first two rows are the limits used when selecting isolated hosts, and the bottom four rows are the values for determining bins for nearby neighbours.  Throughout the rest of the paper, stellar masses refer to the SDSS definitions (first column).}
\label{tab:MF}
\end{center}
\end{table}

As Fig.~\ref{fig:MF} depicts, the \UM{}'s galaxy stellar mass function has more high-mass objects compared to the SDSS MPA-JHU spectroscopic catalogue \citep[\S\ref{sec:obs_spec}; ][]{Kauffmann03, Brinchmann04}. The most significant contribution to this difference is from the treatment of galaxy light profiles as described in \cite{Bernardi13}.  Corrections for these effects are not included in the SDSS stellar mass function (determined in \citealt{Brinchmann04}), but are included in the stellar mass function constraints used in the \UM{} \citep[see appendix C in][]{Behroozi19}.

To account for differences in stellar mass definitions, analogous stellar mass cutoffs in the \UM{} were chosen such that the cumulative number density of galaxies with greater masses matched the cumulative number density expected from the SDSS. Table~\ref{tab:MF} lists equivalent mass values from SDSS MPA-JHU and \UM{} data; in this paper, stellar masses in the text and in figures are values matching SDSS data.

\subsubsection{Density Distribution Normalisation}
\label{sec:density_offset}

A second systematic is the normalisation between the \UM{} and SDSS neighbour density distributions. The \UM{} assumes observed stellar masses have the same biases for both star-forming and quiescent galaxies. In the real Universe, this may not be the case because the differences in metallicity, dust, and star formation histories between quiescent and star-forming galaxies will induce different biases in the inferred stellar masses. For example, the fraction of very old stars is not constrained well from SED-fitting, as old stars contribute very little light.  As a result, the fraction of old stars depends more strongly on the adopted star formation history priors.  This systematic tends to affect quiescent galaxies more strongly, because a greater fraction of their stellar populations were formed at early times \citep{Pacifici16}.  Hence, there can be significant correlations between errors in stellar mass estimates and star formation activity \citep[e.g., private communication,][]{Leja}. In this scenario, our isolated quiescent hosts from the SDSS may have higher stellar masses than the star-forming hosts.
Thus, the true stellar mass distributions will be different for the star-forming and the quiescent galaxies, which implies that the halo masses for the two populations will be systematically different as well. Since a galaxy with a larger halo mass will have a deeper gravitational potential well, it will also be surrounded by more satellite galaxies. The density distribution of nearby neighbours will track the halo mass, leading to a normalisation offset, though the shape of the neighbour density distribution will not be significantly impacted (see Appendix \ref{app:shape_ratio}). We correct for this systematic effect by multiplying by a constant normalisation factor to the \UM{} neighbour density distributions to match the SDSS neighbour density distributions from 1.25 - 2.0 Mpc because this region has the least correlation with accretion rates. Typical offset values are $\lesssim 0.2$ dex. For example, with no correlation between halo and galaxy growth ($\rho = 0.0$), and using neighbours with stellar masses $\log_{10}(M_*/M_\odot) > 9.0$, the star-forming galaxy analogues in the \UM{} require a normalisation factor of $-0.25\pm 0.25$ dex to match the star-forming galaxies in SDSS, and the quiescent galaxy analogues require a normalisation factor of of $0.04 \pm 0.11$ dex to match the quiescent galaxies in the SDSS. The difference between the star-forming and quiescent host stellar mass offsets is consistent with that found in other observational studies (e.g., private communication, \citealt{Leja}, based on the analysis in \citealt{Leja2020}).

As an alternative approach to address the density normalisation offset, we can use isolated hosts with different stellar masses from the \UM{} to match the density distribution normalisations from the SDSS. We take this approach in \S\ref{sec:results_UMoffset}: we compare (1) the neighbour density distributions around isolated hosts with $10.5 < \log_{10}(M_*/M_\odot) < 11.0$, which matches the density distribution normalisation for quiescent SDSS hosts to (2) the neighbour density distributions around isolated hosts with $10.25 < \log_{10}(M_*/M_\odot) < 10.75$, which matches the density distribution normalisation for star-forming SDSS hosts. Comparing these two samples is essentially a test of no correlation ($\rho = 0$), and in this case, we do not require a density distribution normalisation factor.

\subsubsection{Weighting for SDSS Data}
\label{sec:bkgd_weight}

Finally, background and foreground objects represent the main source of noise in the neighbour density distributions for SDSS. The Poisson variance in unassociated source counts is proportional to the expected number of unassociated sources, which is proportional to the on-sky (angular) area of the annulus for counting nearby neighbours.  Hence, we weight the neighbour density distribution for each host by the inverse of the on-sky annulus area $A$,
\begin{equation}
    w_{z} = \frac{1}{A} \propto D_A^2 \appropto z^2 \, ,
    \label{eq:z_weight}
\end{equation}
where $D_A$ is the angular diameter distance, and the last proportionality is valid at low redshifts. This corresponds to inverse-variance weighting, which maximises our signal-to-noise ratio. 

Our analyses account for both this unassociated source weight as well as stellar mass completeness,
\begin{equation}
    w_{\mathrm{tot}} = w_z \cdot w_{*} \, ,
    \label{eq:w_tot}
\end{equation}
where 
\begin{equation}
    w_* = \frac{1}{V_{\mathrm{max}}(M_*)}
    \label{eq:w_vol}
\end{equation}
is the inverse of the co-moving volume $V_{\mathrm{max}}(M_*)$ out to which galaxies at that stellar mass can be observed given SDSS spectroscopic survey limits (\S\ref{sec:obs_spec}).

We calculate the average stacked neighbour density distribution $\bar{n}$ around isolated hosts as
\begin{equation}
    \bar{n} = \frac{ \sum_{i} w_{\mathrm{tot},i} \, (H_i - \bar{R_i})}{\sum_i w_{\mathrm{tot},i}} \, ,
    \label{eq:stack_n}
\end{equation}
where $H_i - \bar{R_i}$ is the background-subtracted neighbour density distribution for each host. $H_i$ is the number density of nearby neighbours around each isolated host and and $\bar{R_i}$ is the average nearby neighbour density around the associated random pointings.  We also calculated $\bar{n}$ using only the stellar mass completeness weights ($w_*$) and confirmed that our results do not change.

\section{Results}
\label{sec:results}

Below, we present our analysis results from applying our technique to the isolated hosts from the SDSS to test whether we observe positive correlations between dark matter accretion and recent star formation activity. In \S\ref{sec:results_no_offset}, we present the shape ratio from the neighbour density distributions in the SDSS. In \S\ref{sec:results_UMoffset}, we account for the possibility of a stellar mass offset between star-forming and quiescent hosts, which may affect the shapes of the distributions, to test the robustness of our findings. Further tests are presented in the appendices.

\subsection{Results Without Stellar Mass Offset}

\label{sec:results_no_offset}

\begin{figure*}
    \centering
    \includegraphics[width=\textwidth]{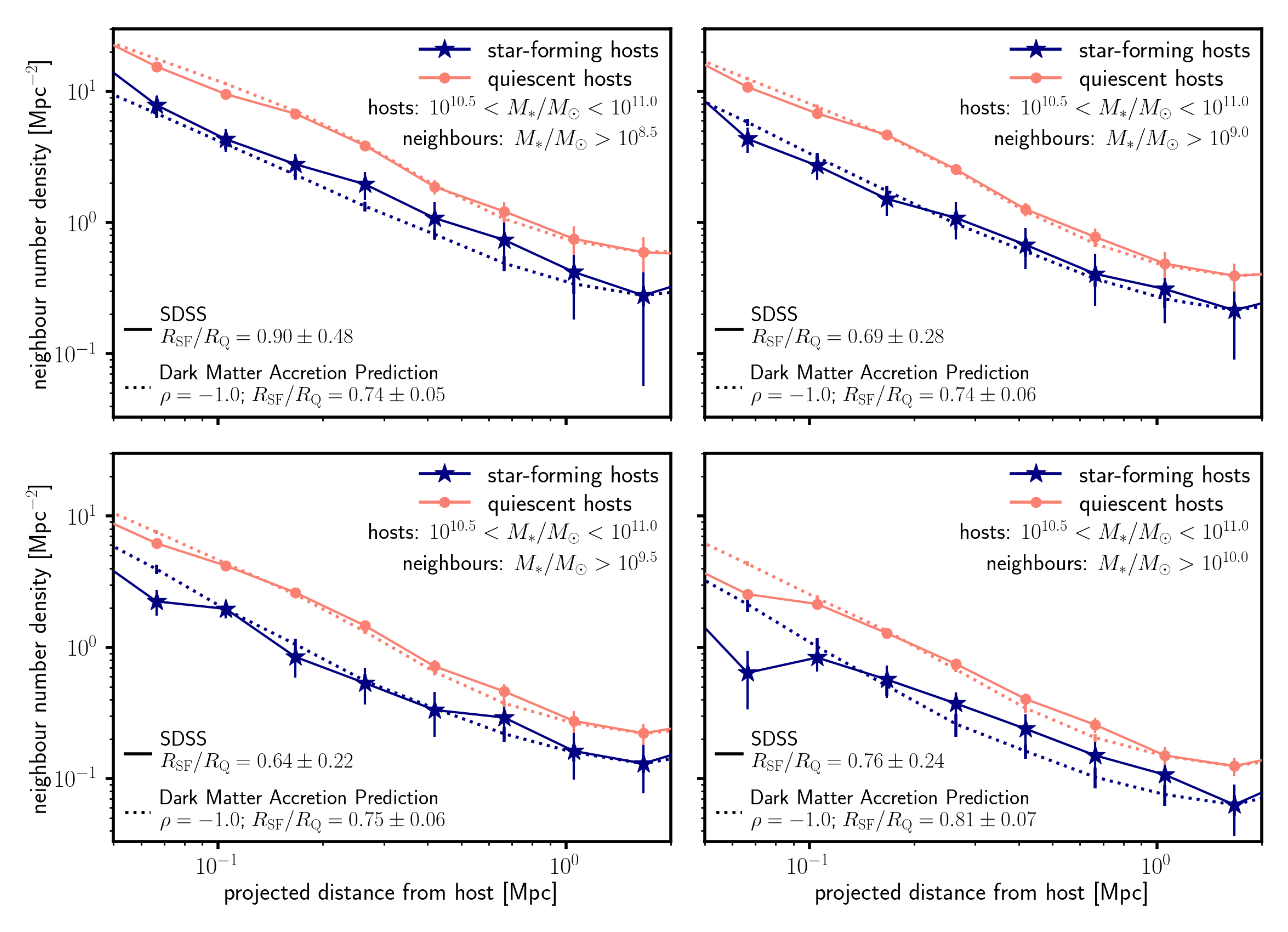}%\\[-4ex]
    \caption{Observed neighbour density distributions for neighbours around isolated hosts in the SDSS are not consistent with positive correlations between halo dark matter accretion rates and recent star formation activity.  Star-forming isolated galaxies have shallower neighbour profiles than quiescent isolated galaxies ($R_\mathrm{SF}/R_\mathrm{Q}<1$).  These results are inconsistent with positive correlations between dark matter accretion rates and SSFRs at typically $\gtrsim 85\%$ confidence (Table \ref{tab:SDSS_results}). The panels compare the neighbour density distributions from SDSS for the four different neighbour galaxy stellar mass thresholds (as indicated by the inset text) versus the best-fitting dark matter accretion prediction, which has a strong anti-correlation ($\rho = -1.0$). As described in \S\ref{sec:density_offset}, a normalisation factor has been included in the plots from the \UM{} to match the observed neighbour density profiles. The redshift distributions of the star-forming and quiescent isolated hosts from the SDSS are similar, and both have an average redshift $z=0.074$.}
    \label{fig:results}
\end{figure*}

By definition, the shape ratio $R_\mathrm{SF}/R_\mathrm{Q}$ should be $1.0$ in the case of no correlation ($\rho = 0$) because the shapes of the neighbour density distributions around star-forming and quiescent hosts should be the same. However, as shown in Fig.~\ref{fig:results} and Table \ref{tab:SDSS_results}, isolated star-forming galaxies have shallower neighbour distributions than isolated quiescent galaxies ($R_\mathrm{SF}/R_\mathrm{Q}<1$).  This statement is independent of the neighbour mass threshold (Fig.\ \ref{fig:results}), and implies that positive correlations between dark matter halo accretion and galaxy star formation activity are ruled out with $\gtrsim 85$\% confidence (i.e., $\rho\le 0$, Table \ref{tab:SDSS_results}). Indeed, for neighbours with $\log_{10}(M_*/M_\odot) > 9.0$ around isolated hosts in the SDSS, the observed results are most consistent with theoretical predictions for a dark matter accretion correlation rate that is fully anticorrelated with recent galaxy star formation activity ($\rho = -1.0$).  However, given the uncertainties, weakly negative and/or zero correlations are still plausibly consistent with observations.

We have tested many variations on the method presented in \S \ref{sec:methods}, and find in all cases a strong observational preference for $\rho \le 0$ (i.e., $R_\mathrm{SF}/R_\mathrm{Q}<1$).  Our results do not change significantly if either: (1) we exclude fibre-collided galaxies with masses from the NYU-VAGC, (2) we only weight by stellar mass completeness and exclude the inverse variance weights (Eqs.~\ref{eq:z_weight}-\ref{eq:stack_n}), (3) we include redder photometric sources up to $g-r < 1.25$ (\S\ref{sec:obs_phot}), or (4) we select neighbours based on luminosities (Appendix \ref{app:Ms_Mr_bins}) or the stellar mass proxy from \cite{Bell03} (Appendix \ref{app:Ms_gr_fit_Bell}).

We also find that for the most massive nearby neighbours ($\log(M_\ast/M_\odot) \gtrsim 9.5$), we observe a deficit close to the isolated hosts (projected distances $R< 0.125$ Mpc) as compared to theoretical predictions for both star-forming and quiescent hosts. This deficit could indicate either a reduced efficiency in detecting these neighbours and/or short disruption timescales.  To make sure that these potential effects do not  bias our conclusions, we repeated our analysis on the neighbour density distributions excluding neighbours within 0.125 Mpc and found that this does not significantly change our results. Finally, we replicated our analysis excluding the \UM{} orphan model (\S~\ref{sec:UM}), but we found that it could not fully explain the observed deficit (Appendix \ref{app:orphan}).

\begin{table}
\begin{center}
\begin{tabular}{c|c|c}
    Neighbour $M_*$ Selection & \multirow{2}{*}{$R_\mathrm{SF}/R_\mathrm{Q}$} & Confidence Level\\
    $\log_{10}(M_*/M_\odot)$ & & $\rho \leq 0.0$\\
    \hline 
    %%% g-r < 1.0 values
    8.50 & $0.899 \pm 0.480$ & 58.32\% \\
    9.00 & $0.689 \pm 0.284$ & 86.34\% \\
    9.50 & $0.638 \pm 0.216$ & 95.35\% \\
    10.00 & $0.761 \pm 0.237$ & 84.29\% \\
    \hline
\end{tabular}
\caption{The shape ratios for neighbour density distributions around isolated hosts in the SDSS are most consistent with an anti-correlation between dark matter accretion and SSFRs (i.e., $R_\mathrm{SF}/R_\mathrm{Q} < 1$ and thus $\rho < 0$). In this table and Fig.~\ref{fig:results}, we include values for all neighbour stellar mass selection limits as determined from their $g-r$ colours (\S\ref{sec:obs_phot}). To determine the confidence levels for non-positive correlation strengths, we compare the observed shape ratios to $R_\mathrm{SF}/R_\mathrm{Q} \equiv 1.0$, which is what we should measure by definition for $\rho=0$.}
\label{tab:SDSS_results}
\end{center}
\end{table}

\subsection{Results Assuming a Stellar Mass Offset in \UM{}}
\label{sec:results_UMoffset}

\begin{figure*}
    \centering
    \includegraphics[width=\textwidth]{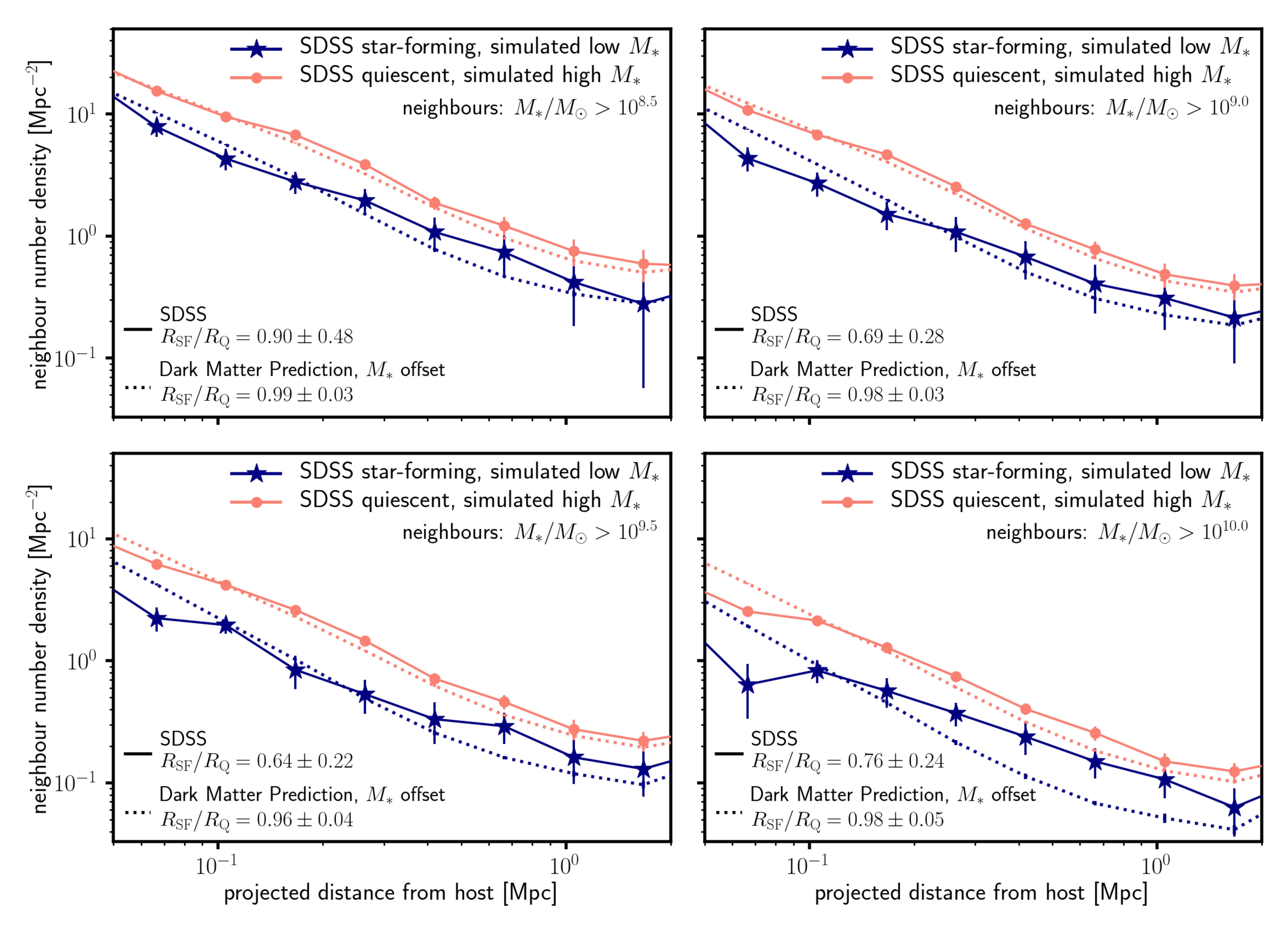}%\\[-4ex]
    \caption{By matching the neighbour density distribution normalisations from the \UM{} to those osberved in the SDSS, we still find that our SDSS data is consistent with correlation strengths $\rho \leq 0$. To test the impact of the stellar mass offset in the isolated host samples (\S\ref{sec:density_offset}), we select isolated hosts from the \UM{} for a (1) `high $M_*$ isolated host sample' with $10.5 < \log_{10}(M_*/M_\odot) < 11.0$ to correspond to the SDSS quiescent hosts and (2) a `low $M_*$ isolated host sample' with $10.25 < \log_{10}(M_*/M_\odot) < 10.75$ to correspond to the SDSS star-forming hosts. The normalisations of the corresponding neighbour density distributions are much more similar with this sampling approach. The \UM{} samples are effectively a prediction for $\rho = 0$, and with these isolated host samples, the predicted shape ratio is slightly less than 1.0 ($R_\mathrm{SF}/R_\mathrm{Q} \sim 0.9$). However, the neighbour density distributions from the SDSS are still shallower than the \UM{} predictions, which is consistent with our finding that the SDSS isolated hosts show no evidence for positive correlations between star formation and dark matter accretion.}
    \label{fig:results_Ms_offset}
\end{figure*}

As described in \S~\ref{sec:density_offset}, the density distribution normalisation offset between the predicted neighbour density distributions in the \UM{} versus those observed in the SDSS is due to the fact that the \UM{} assumes the same biases in observed stellar masses for both star-forming and quiescent galaxies. However, this may not be true in the real Universe, and so there may be an offset in the stellar masses reported in the SDSS for star-forming versus quiescent galaxies. While we do not expect this to affect the shapes of the neighbour density distributions, it will impact the normalisations. As another strategy to account for this effect, we measured the density distributions around different isolated host stellar mass ranges in the \UM{} to match the observed neighbour density normalisations. Specifically, we selected a ``high $M_*$ isolated host sample'' in the \UM{} with $10.5 < \log_{10}(M_*/M_\odot) < 11.0$ to correspond to the SDSS quiescent hosts, and we selected a ``low $M_*$ isolated host sample'' in the \UM{} with $10.25 < \log_{10}(M_*/M_\odot) < 10.75$ to correspond to the SDSS star-forming hosts. By comparing the neighbour density distributions around the isolated hosts in these two different stellar mass bins, we effectively construct a prediction for $\rho=0$ that accounts for the stellar mass offset in the isolated host population (Fig.~\ref{fig:results_Ms_offset}). 

With the different isolated host stellar masses, the predicted shape ratio for $\rho = 0$ is slightly less than 1.0 ($R_\mathrm{SF}/R_\mathrm{Q} \sim 0.9$). To calculate the significance of our results with the stellar mass offset included in the \UM{} data, we performed an integral over the distribution of shape ratios from our jackknife sampling
%\begin{equation}
%    C = \int f_\mathrm{obs}\left(\leq %\frac{R_\mathrm{SF}}{R_\mathrm{Q}}\right) P_\mathrm{m}\left(\frac{R_\mathrm{SF}}{R_\mathrm{Q}}\right) \, d\frac{R_\mathrm{SF}}{R_\mathrm{Q}}
%    \label{eq:confidence}
%\end{equation}

\begin{eqnarray}
    C & =& \int f_\mathrm{obs}\left(\leq R_\mathrm{obs}'\right) P_\mathrm{m}\left(R_\mathrm{m}'\right) \, d R_\mathrm{m}'
    \label{eq:confidence}\\
    R_\mathrm{obs}' &=& \frac{1}{\sqrt{n_\mathrm{jk, obs}-1}} \left(\frac{R_\mathrm{SF}}{R_\mathrm{Q}} - \left\langle\frac{R_\mathrm{SF}}{R_\mathrm{Q}}\right\rangle_\mathrm{obs}  \right) + \left\langle\frac{R_\mathrm{SF}}{R_\mathrm{Q}}\right\rangle_\mathrm{obs} \\
     R_\mathrm{m}' &=& \frac{1}{\sqrt{n_\mathrm{jk,m}-1}} \left(\frac{R_\mathrm{SF}}{R_\mathrm{Q}} - \left\langle\frac{R_\mathrm{SF}}{R_\mathrm{Q}}\right\rangle_\mathrm{m}  \right) + \left\langle\frac{R_\mathrm{SF}}{R_\mathrm{Q}}\right\rangle_\mathrm{m}
\end{eqnarray}
where $C$ is the confidence level that $\rho \leq 0$, $R'$ is the distribution of $R_\mathrm{SF}/R_\mathrm{Q}$ from the SDSS (``obs'') or \UM{} (model ``m'') jackknife samples (including the correction for the variance of the jackknife samples being a factor $(n-1)$ smaller than the expected sample variance),
$f_\mathrm{obs}$ is the fraction of jackknife samples from the SDSS that return a shape ratio less than $R_\mathrm{SF}/R_\mathrm{Q}$, and $P_\mathrm{m}$ is the probability of that value of $R_\mathrm{SF}/R_\mathrm{Q}$ from the \UM{}. We calculated this integral both using the entire simulation sample of 14 snapshots. In addition, we also analyzed a single snapshot with $a=0.956$ using SMDPL. Since the SMDPL has a larger co-moving volume than \textit{Bolshoi-Planck} (400 Mpc$/h$ per side versus 250 Mpc$/h$ per side, respectively), using SMDPL for this analysis reduces our uncertainties. The results are detailed in  Table \ref{tab:results_UM_offset}.
The results are detailed in  Table \ref{tab:results_UM_offset}.

\begin{table}
\begin{center}
\begin{tabular}{c|c|c}
    Neighbour $M_*$ Selection & \multicolumn{2}{c}{Confidence Level $\rho \leq 0$}\\%\multirow{2}{*}{$R_\mathrm{SF}/R_\mathrm{Q}$} & Confidence Level\\
    $\log_{10}(M_*/M_\odot)$ & full \UM{} & $a=0.956$ \\%& $\rho \leq 0.0$\\
    \hline 
    8.50 &  59.24\% &  58.75\% \\
    9.00 & 77.43\% & 80.04\% \\
    9.50 & 90.54\% & 90.93\% \\
    10.00 &  72.96\% & 78.00\% \\
    \hline
\end{tabular}
\caption{The shape ratios for neighbour density distributions around isolated hosts in the SDSS are most consistent with an anti-correlation between dark matter accretion and SSFRs (i.e., $R_\mathrm{SF}/R_\mathrm{Q} < 1$ and thus $\rho < 0$), even when adjusting the \UM{} isolated host samples to account for the stellar mass offset between the \UM{} and the SDSS (\S\ref{sec:density_offset}). For this table, the \UM{} sample compared  (1) a ``high $M_*$ isolated host sample'' with $10.5 < \log_{10}(M_*/M_\odot) < 11.0$ to correspond to the SDSS quiescent hosts and (2) a ``low $M_*$ isolated host sample'' with $10.25 < \log_{10}(M_*/M_\odot) < 10.75$ to correspond to the SDSS star-forming hosts. This effectively tests $\rho = 0.0$, and we calculated the confidence levels that the SDSS data shows $\rho \leq 0$ using Eq.~(\ref{eq:confidence}). The second column is the confidence level using all of the \UM{} data with haloes from \textit{Bolshoi-Planck} (i.e., all 14 snapshots), and the right column analyzes the data using only one snapshot ($a=0.956$) with haloes from SMDPL. With the single snapshot ($a=0.949$), our ability to constrain the correlation strength with a neighbour selection limit $M_* > 10^{10.0} M_\odot$ is limited because there are too few neighbours that pass this threshhold. Similarly, we are also limited in constraining power for the lowest-mass neighbour selection limit $M_* > 10^{8.5} M_\odot$ because we cannot observe neighbours down to those masses throughout the entire isolated host redshift range.}
\label{tab:results_UM_offset}
\end{center}
\end{table}

\begin{figure}
    \centering
    \includegraphics[width=\columnwidth]{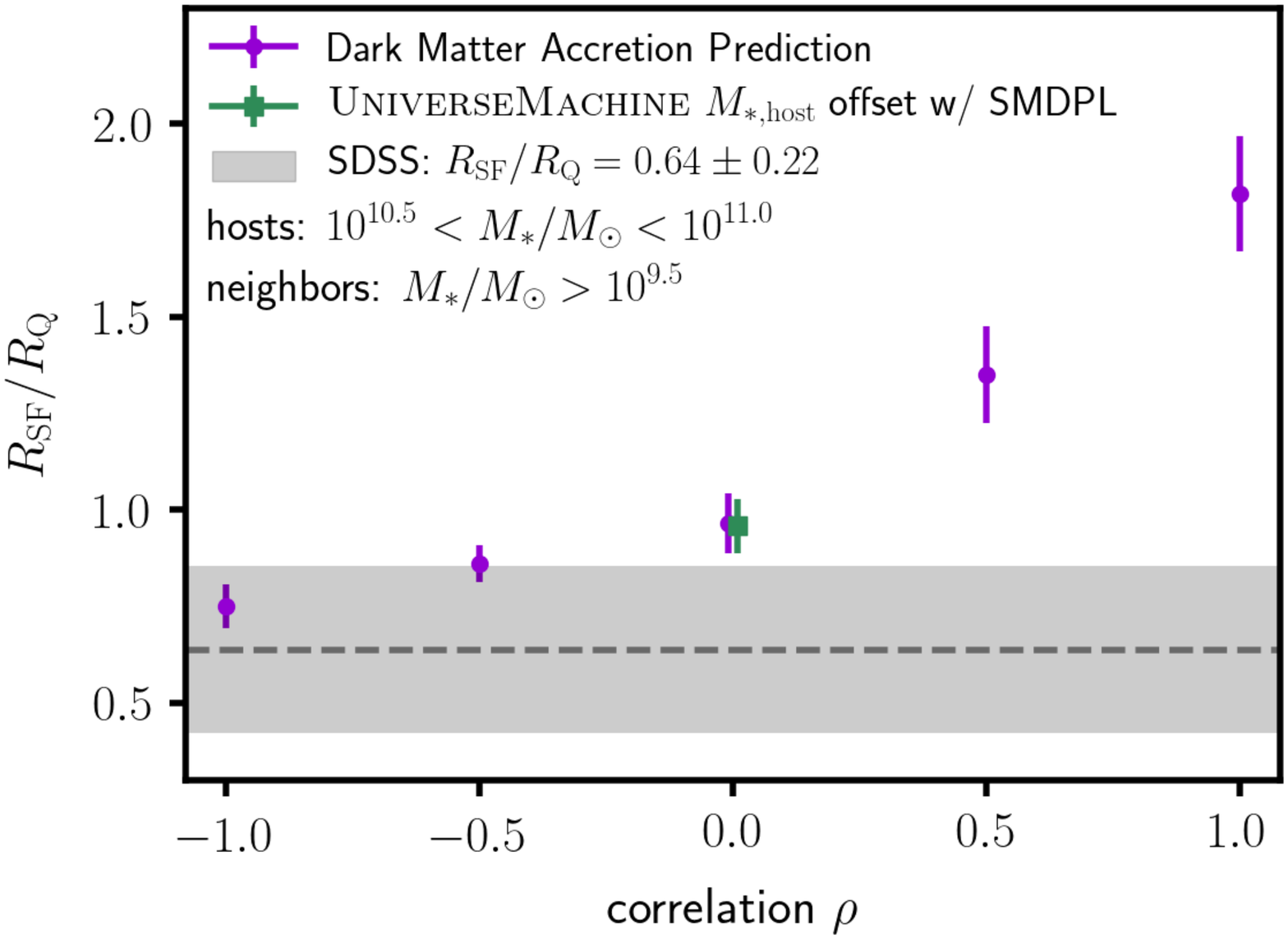}\\[-2ex]
    \caption{Our results from the SDSS show no evidence for positive correlations between dark matter accretion and recent star formation activity. To emphasize this finding, we plot the predicted shape ratios from the \UM{} for correlation strengths from $\rho = -1.0$ to $\rho=1.0$ (purple circles), as well as the prediction from using different stellar mass selections for the isolated hosts to match the observed neighbour density distribution normalisations (effectively $\rho = 0.0$; green square). The grey shaded region is the shape ratio from the SDSS data, which is shallower than any predictions for $\rho > 0$. This particular plot uses data for the neighbour selection limit $M_* > 10^{9.5} M_\odot$, and we find similar results for the other neighbour selection limits (see Appendix \ref{app:shape_ratio_plots}). A slight offset in the $x$-position has been introduced for the points representing \UM{} preductions for $\rho=0$ for plot readability.}
    \label{fig:shape_ratio_results_Ms_offset_9.5}
\end{figure}

Finally, to emphasize that our observed shape ratios are consistent with $\rho \leq 0$, Fig.~\ref{fig:shape_ratio_results_Ms_offset_9.5} compares the predicted shape ratio as a function of $\rho$ from the \UM{} versus the results from the SDSS for neighbours with $M_* > 10^{9.5} M_\odot$. Similar plots for the other neighbour mass selection limits are presented in Appendix \ref{app:shape_ratio_plots}.

\section{Discussion \& Conclusions}
\label{sec:disc_conc}

In this paper, we present a method to observationally constrain the correlation between dark matter accretion and star formation, validate the method on simulated galaxies from the \UM{}, and apply the method to Milky Way-mass galaxies in the SDSS. The method is based on the density distributions of neighbouring galaxies, which we obtain by background subtraction. As a halo accretes more dark matter, we expect that the neighbour density distribution will steepen near the halo, and we confirm this trend in the predicted neighbour density distributions from the \UM{} (Fig.~\ref{fig:accret_rate}). We quantify this effect using a shape ratio optimised to be sensitive to differences in the distributions around high-accreting versus low-accreting hosts. Using a sample of isolated star-forming and quiescent (as determined by SSFRs) host galaxies from the SDSS with $10.5 < \log_{10}(M_*/M_\odot) < 11.0$, our results suggest that a positive correlation between dark matter accretion and galaxy SSFRs is ruled out with $\gtrsim 85\%$ confidence.

We note several factors that could affect the interpretation of our results. First, weak correlations might be expected if changes in host galaxy SSFRs happen on timescales much shorter than satellite galaxy periods ($\sim 2 \, t_\mathrm{dyn} \sim 4 \, \mathrm{Gyr}$). Second, our analysis uses the distributions of neighbouring galaxies, but neighbouring galaxies may be subject to effects that are not adequately modeled in dark matter simulations. For example, dynamical friction and tidal disruption act to reduce the number of satellites. If these effects occur at different rates between star-forming and quiescent host galaxies and are not adequately captured in our simulations, then our measured shape ratios could be affected. Both of these factors can be addressed by further measurements. O'Donnell et al. (in prep.) will expand the analysis to larger isolated host mass ranges as well as classify hosts by their 4000\r{A} break strength ($D_\mathrm{n}4000$), which is a longer-term diagnostic of a galaxy's star formation.  Additionally, weak lensing techniques can be used to more accurately measure the dark matter mass profile rather than relying on the density profile of neighbouring galaxies.

Our results are consistent with galaxy formation models that are not correlated with fresh accretion at $z=0$. These models instead invoke modest recycling timescales for gas that is ejected from the galaxy, resulting in the gas cooling and reaccreting onto the galaxy \citep[e.g.,][]{Muratov15,vandeVoort16, Keres05, Nelson13, Nelson15, Dekel06}. This process can generate subsequent galaxy growth even in the absence of new accretion onto the host halo. Because only $\sim 20-30\%$ of gas in the host haloes turns into stars for our isolated galaxy sample \citep{Behroozi19}, plenty of gas remains that could support star formation after accretion stops.

Furthermore, our results are consistent with other studies that do not find strong correlations between halo growth and star formation. \cite{Tinker17} studied the fraction of quenched central galaxies (as determined by $D_{n}4000$) in galaxy groups from SDSS data. They found no correlation for $M_* \lesssim 10^{10} M_\odot / h^2$, and at higher masses, they found a only $\sim$5\% increase in the quenched fraction from low to high densities for fixed stellar mass. Similarly, \cite{Behroozi15} also did not find a strong effect of major halo mergers (as probed by close galaxy pairs) on star-formation rates.

Other theoretical studies have found or assumed strong positive correlations between dark matter accretion and star formation, including \cite{WetzelNagai15}, \cite{Becker15}, \cite{Cohn17}, \cite{Rodriguez-Puebla16}, and \cite{Moster18}.  \cite{Behroozi19} measured a strong correlation roughly equivalent to $\rho = 0.6$. However, the \cite{Behroozi19} constraints were primarily driven by satellite galaxies, which have low or negative dark matter accretion rates. The \UM{} did not independently constrain accretion correlations for central/isolated haloes (see \S{}5.11 of \citealt{Behroozi19}). Despite the findings in this paper, we note that most results of these models (e.g., stellar vs. halo mass relations) are likely still valid, as an equivalent model that assumed anti-correlation between accretion rates for isolated central galaxies and their star formation rates would have almost no observational differences except for the measurement in our paper.

Finally, we note that our shape ratio was maximised with $r_\mathrm{split} \equiv 0.316$ Mpc, which is fairly close to the host galaxies' halo virial radii ($\sim 0.05^\circ$ at the median host redshift of $z=0.079$). Future surveys, such as the Dark Energy Spectroscopic Instrument (DESI) Survey\footnote{\url{https://www.desi.lbl.gov/}} \citep{DESI_1science}, will detect larger numbers of such galaxies at higher redshifts and thus will provide stronger constraints on the correlation strength between dark matter accretion and star formation. Additionally, a larger number of isolated galaxies will enable analyses that measure the correlation between dark matter accretion and other host properties, including metallicity, velocity dispersion, and AGN activity.

\section*{Acknowledgements}
We thank Gurtina Besla, Marla Geha, Elisabeth Krause, Dan Marrone, and Eduardo Rozo for helpful comments during the preparation of this paper.

Support for this research came partially via program number HST-AR-15631.001-A, provided through a grant from the Space Telescope Science Institute under NASA contract NAS5-26555.  PB was partially funded by a Packard Fellowship, Grant \#2019-69646.  An allocation of computer time from the UA Research Computing High Performance Computing (HPC) at the University of Arizona is gratefully acknowledged. The coding and plots created for this work were done with Python packages NumPy \citep{numpy, numpy_array} and Matplotlib \citep{matplotlib}.

The Bolshoi simulations have been performed within the Bolshoi project of the University of California High-Performance AstroComputing Center (UC-HiPACC) and were run at the NASA Ames Research Center. Funding for the Sloan Digital Sky Survey IV has been provided by the Alfred P. Sloan Foundation, the U.S. Department of Energy Office of Science, and the Participating Institutions. SDSS-IV acknowledges
support and resources from the Center for High-Performance Computing at
the University of Utah. The SDSS web site is www.sdss.org.
SDSS-IV is managed by the Astrophysical Research Consortium for the 
Participating Institutions of the SDSS Collaboration including the 
Brazilian Participation Group, the Carnegie Institution for Science, 
Carnegie Mellon University, the Chilean Participation Group, the French Participation Group, Harvard-Smithsonian Center for Astrophysics, 
Instituto de Astrof\'isica de Canarias, The Johns Hopkins University, Kavli Institute for the Physics and Mathematics of the Universe (IPMU) / 
University of Tokyo, the Korean Participation Group, Lawrence Berkeley National Laboratory, 
Leibniz Institut f\"ur Astrophysik Potsdam (AIP),  
Max-Planck-Institut f\"ur Astronomie (MPIA Heidelberg), 
Max-Planck-Institut f\"ur Astrophysik (MPA Garching), 
Max-Planck-Institut f\"ur Extraterrestrische Physik (MPE), 
National Astronomical Observatories of China, New Mexico State University, 
New York University, University of Notre Dame, 
Observat\'ario Nacional / MCTI, The Ohio State University, 
Pennsylvania State University, Shanghai Astronomical Observatory, 
United Kingdom Participation Group,
Universidad Nacional Aut\'onoma de M\'exico, University of Arizona, 
University of Colorado Boulder, University of Oxford, University of Portsmouth, 
University of Utah, University of Virginia, University of Washington, University of Wisconsin, 
Vanderbilt University, and Yale University.

\section*{Data Availability}
No new data were generated or analysed in support of this research. The isolated host catalogs derived from SDSS DR16 \cite{SDSS_DR16} and from the \UM{} \cite{Behroozi19} are available at \url{https://github.com/caodonnell/DM_accretion}.

\bibliographystyle{mnras}
\bibliography{paper.bib}
\bsp	% typesetting comment

\appendix
\section{Neighbour Density Distributions Using Different Neighbour Selections}
\label{sec:appendix}

In our analysis, we binned neighbouring galaxies based on their stellar masses as determined by their $g-r$ colours because stellar masses are expected to be more robust throughout a satellite galaxy's orbit. However, previous studies have binned  neighbours by their luminosities \citep[e.g.,][]{More16, Baxter17}, and our relation between $g-r$ colours and mass-to-light ratios differed from the relation found in \cite{Bell03} because of differences in assumptions used. Here, we explore implications of our analysis choices.

\subsection{Luminosity versus Stellar Mass Binning}
\label{app:Ms_Mr_bins}

The left panel of Fig.~\ref{fig:Mr_Bell} compares the neighbour density distributions around isolated hosts with a neighbour selection of $M_*/M_\odot > 10^{9.0}$ versus a neighbour selection of $M_r < -18.0$. Based on the SDSS DR16 spectroscopic galaxies used to develop our stellar mass proxy (\S\ref{sec:obs_phot}), we found that $\gtrsim 90\%$ of galaxies with $\log_{10}(M_*/M_\odot) > 9.0$  were brighter than $M_r < -18.0$. The two sets of neighbour density distributions are very similar, though the luminosity-based threshold identifies more neighbours around star-forming hosts (especially at smaller distances) whereas the distributions around quiescent hosts are similar.  This results in a slightly higher shape ratio $R_{\mathrm{SF}}/R_\mathrm{Q}$, though they are not statistically different ($0.69\pm 0.28$ and $0.78 \pm 0.22$ for stellar mass and luminosity selections, respectively). This difference is consistent with our findings that close to isolated hosts, neighbour galaxies tend to be bluer, but at larger distances, the colours of neighbour galaxies are more similar (Fig.~\ref{fig:Mstellar_gr_hist}).

\begin{figure*}
    \centering
    \includegraphics[width=\textwidth]{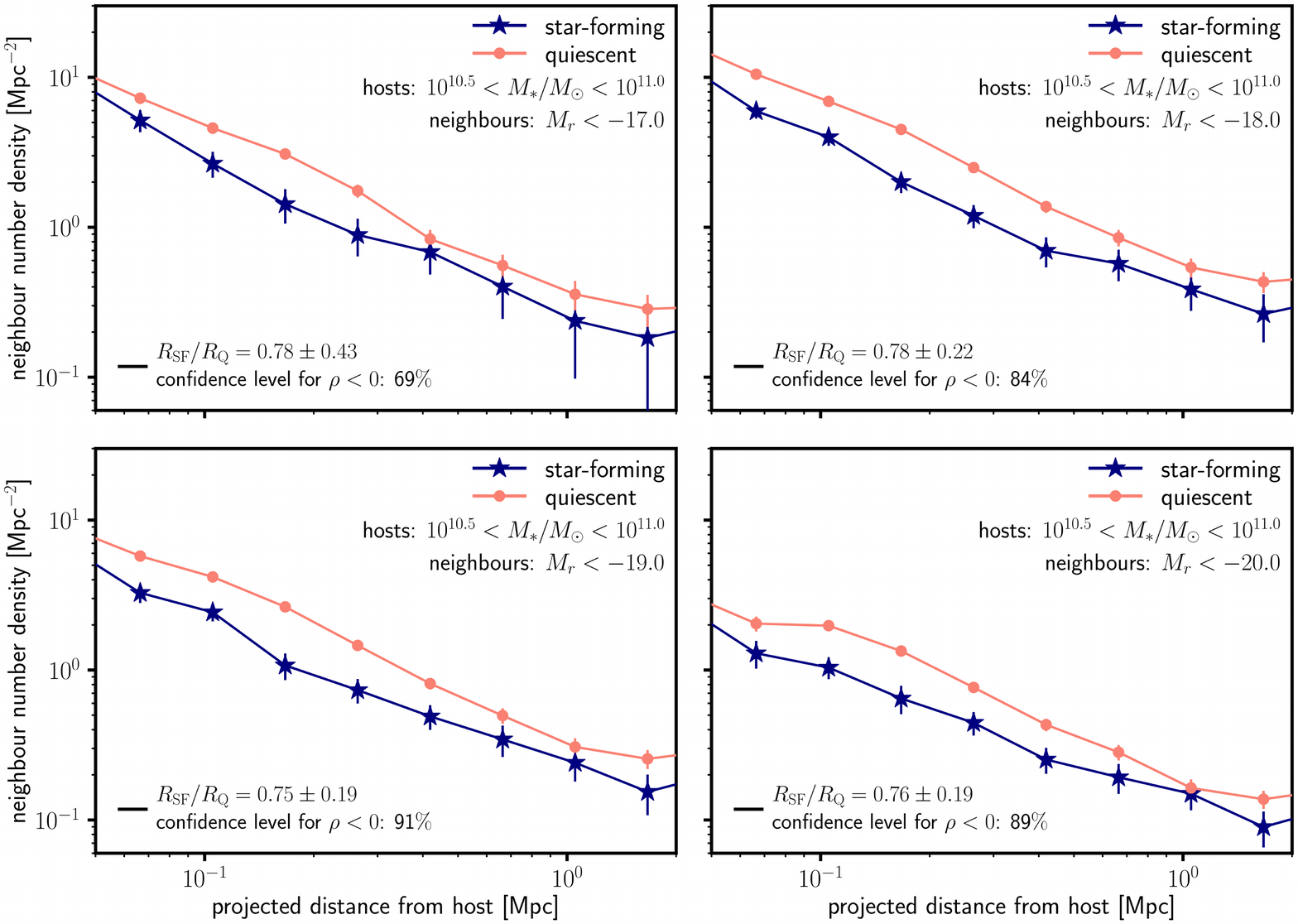}\\[-5ex]
    \caption{Neighbour density distributions around isolated hosts are still consistent with anti-correlation between dark matter accretion and star formation when selecting neighbours by luminosity. These panels compare the neighbour density distributions around star-forming versus quiescent isolated hosts from the SDSS for four different luminosity selections. Positive correlations are ruled out with typically $\gtrsim 85\%$ confidence as indicated in the inset text.}
    \label{fig:results_Mr}
\end{figure*}

We also present neighbour density distributions for neighbour selection limits of $M_r < -17.0$ to $M_r < -20.0$ (Fig.~\ref{fig:results_Mr}). These results are also consistent with correlation strengths $\rho \leq 0$ at $\gtrsim 85\%$ confidence.

\subsection{Relation between \texorpdfstring{$M_*$}{Stellar Mass} and \texorpdfstring{$g-r$}{Colour}}
\label{app:Ms_gr_fit_Bell}

\begin{figure*}
    \centering
    \includegraphics[width=0.99\columnwidth]{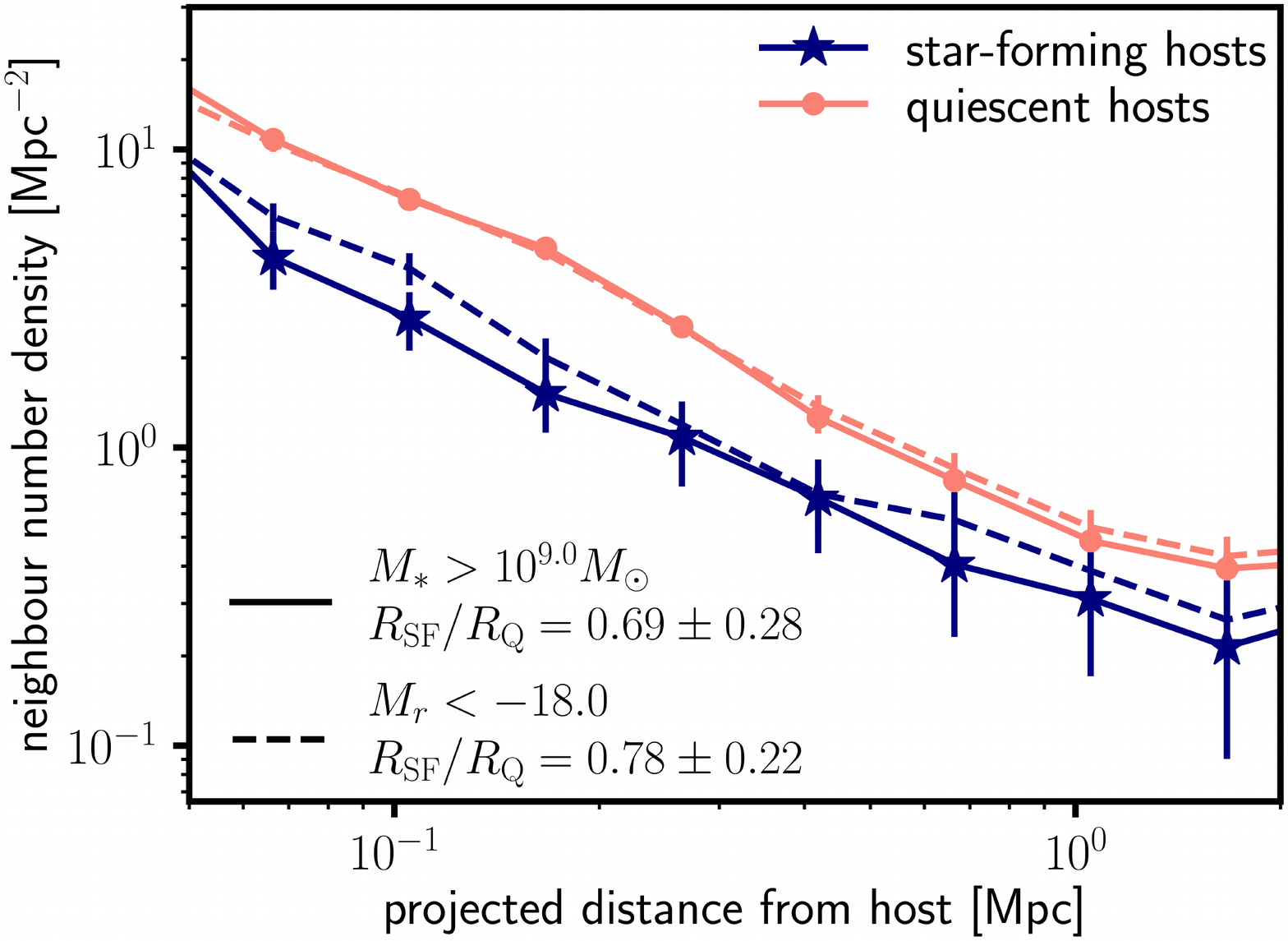}
    \hfill
    \includegraphics[width=0.99\columnwidth]{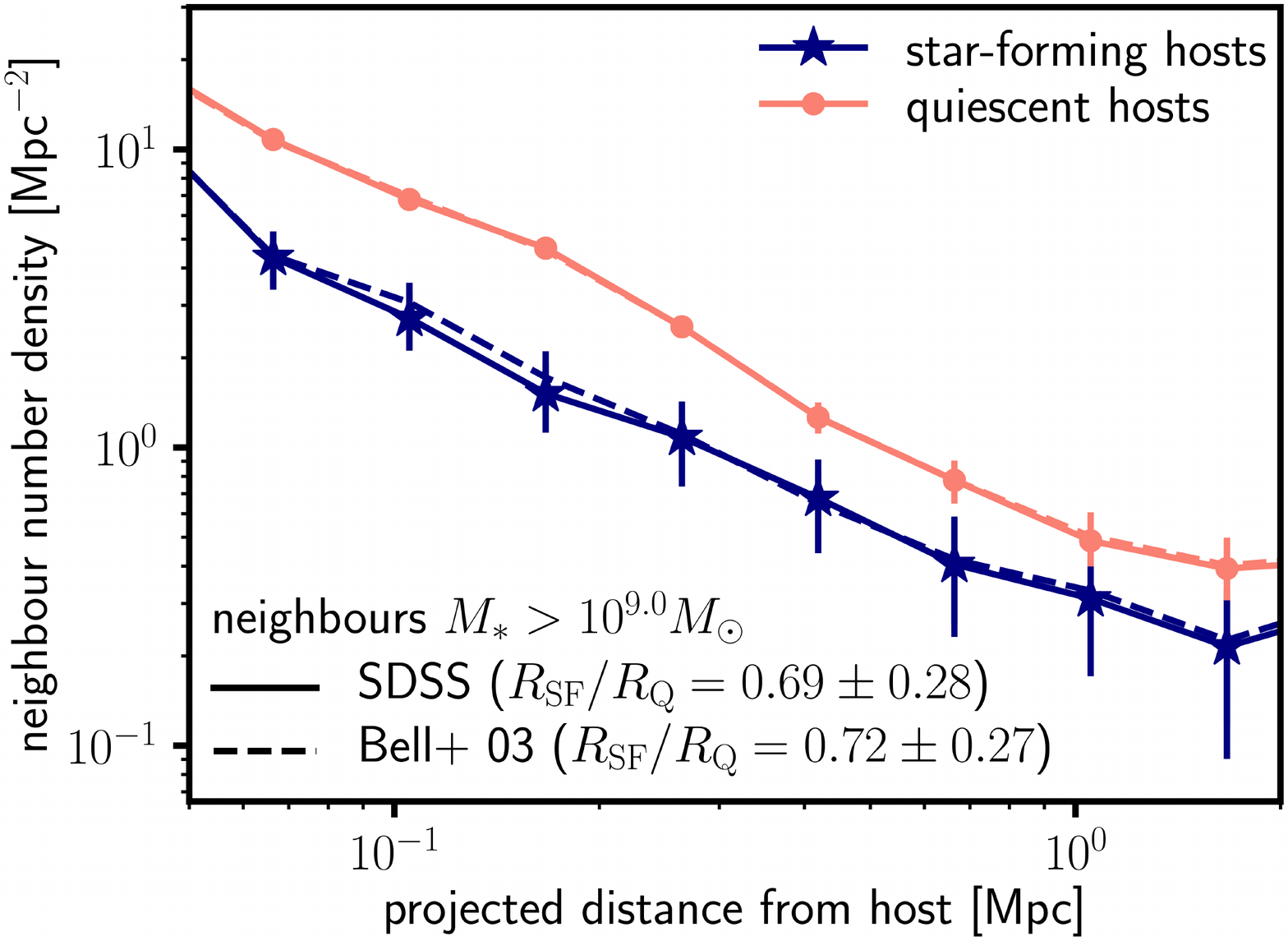}%\\[-1ex]
    \vspace{-\baselineskip}
    \caption{\textbf{Left Panel:} Using a luminosity selection for nearby neighbours versus a stellar mass selection slightly increases the shape parameter, but the difference is not statistically significant. In our analysis, we use stellar mass bins because we expect that the stellar masses of satellite galaxies are more robust throughout their orbits. \textbf{Right Panel:} The differences between our stellar mass proxy and the fit from \protect\cite{Bell03} does not lead to any significant differences in the neighbour density distributions. In our analysis, we use the fit derived from galaxies in the SDSS spectroscopic catalogue with redshifts and stellar masses that match the values used in our nearby neighbour selection. The \protect\cite{Bell03} results above account for the differences in choice of IMF following \protect\cite{Salim07}. Their fit used galaxies over a larger redshift range and $k$-corrected colours, whereas we include galaxies over a smaller redshift range and use dereddened $g-r$ colours. However, these differences do not create appreciable changes in the neighbour density distributions.}
    \label{fig:Mr_Bell}
\end{figure*}

As described in \S\ref{sec:obs_phot}, we fit a relation between mass-to-light ratios for galaxies for the SDSS DR16 spectroscopic catalogues following the approach in \cite{Bell03}. However, our fit differed from the results in \cite{Bell03} even after converting the \cite{Bell03} fit to account for the choice of IMF. This difference is due to the colours used (dereddened versus $k$-corrected) and redshift range of galaxies included in the fit. The right panel of Fig.~\ref{fig:Mr_Bell} compares the neighbour density distribution according to the two fits. We find that the different fits create similar neighbour density distributions with nearly identical shape ratios.

\section{\UM{} with Specific Halo Accretion Rates over \texorpdfstring{$2t_\mathrm{dyn}$}{Longer Timescales} }
\label{app:2tdyn}

\begin{figure*}
    \centering
    \includegraphics[width=\textwidth]{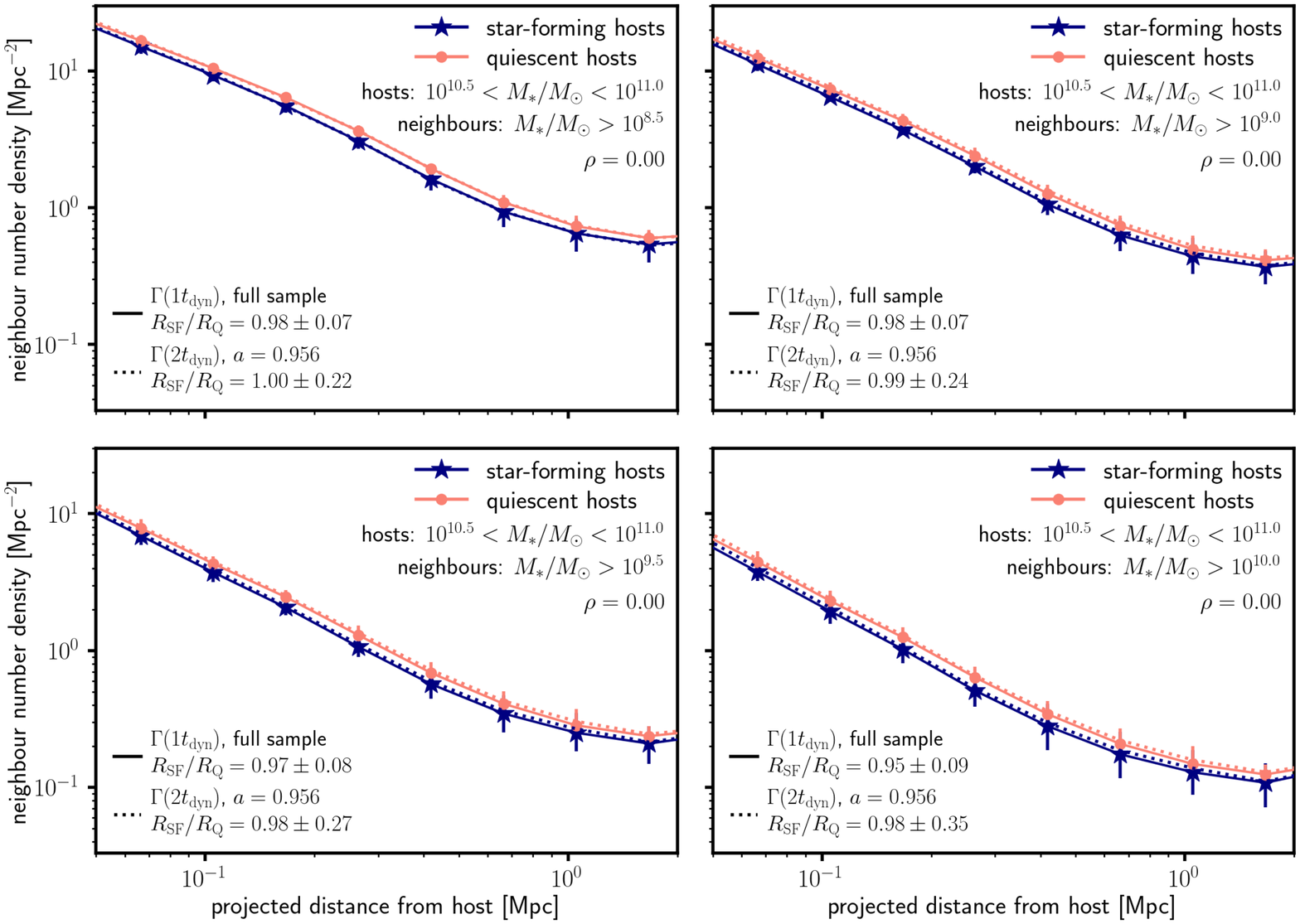}%\\[-2ex]
    \caption{As expected, for no correlation ($\rho=0$) between star formation and dark matter accretion, the shape ratios for the neighbour density distributions around isolated hosts in the \UM{} with$10.5 < \log_{10}(M_*/M_\odot) < 11.0$ is 1.0 (within model scatter). This holds when binning isolated hosts into star-forming versus quiescent bins based on their specific halo accretion rates over the past $t_\mathrm{dyn}$ (solid lines) or past $2t_\mathrm{dyn}$ (dotted lines).}
    \label{fig:2tdyn_rho_0}
\end{figure*}

\begin{figure*}
    \centering
    \includegraphics[width=\textwidth]{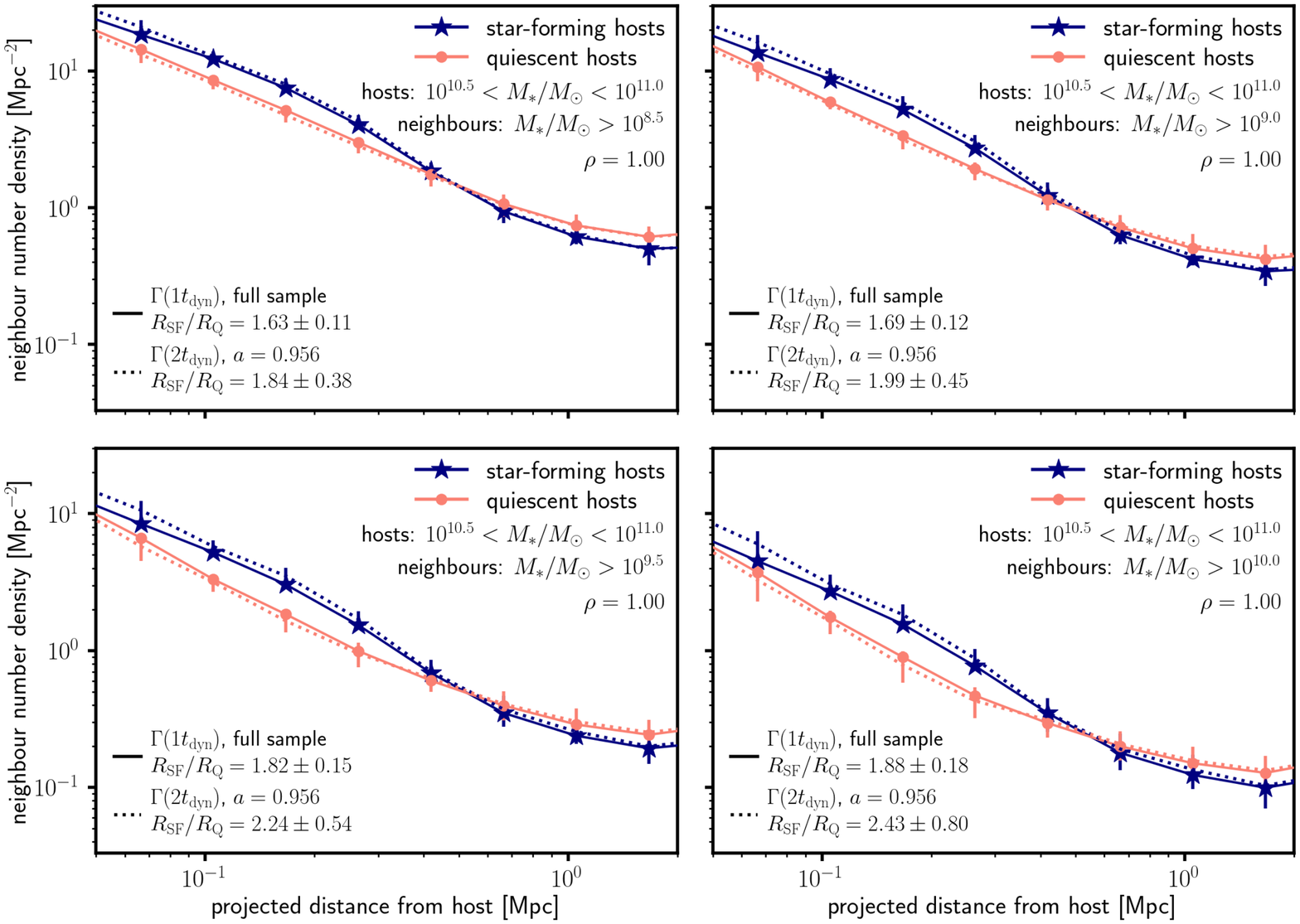}%\\[-2ex]
    \caption{When binning isolated hosts from the \UM{} into star-forming versus quiescent bins based on their specific halo accretion rates over the past $2t_\mathrm{dyn}$ (dotted lines) instead of their past $t_\mathrm{dyn}$ (solid lines), the predicted shape ratios for $\rho=1.0$ are much higher than 1 (corresponding to $\rho = 0.0)$. Based on this finding, we would expect to see larger shape ratios for correlation between dark matter accretion and star formation over longer timescales.}
    \label{fig:2tdyn_rho_100}
\end{figure*}

\begin{figure*}
    \centering
    \includegraphics[width=\textwidth]{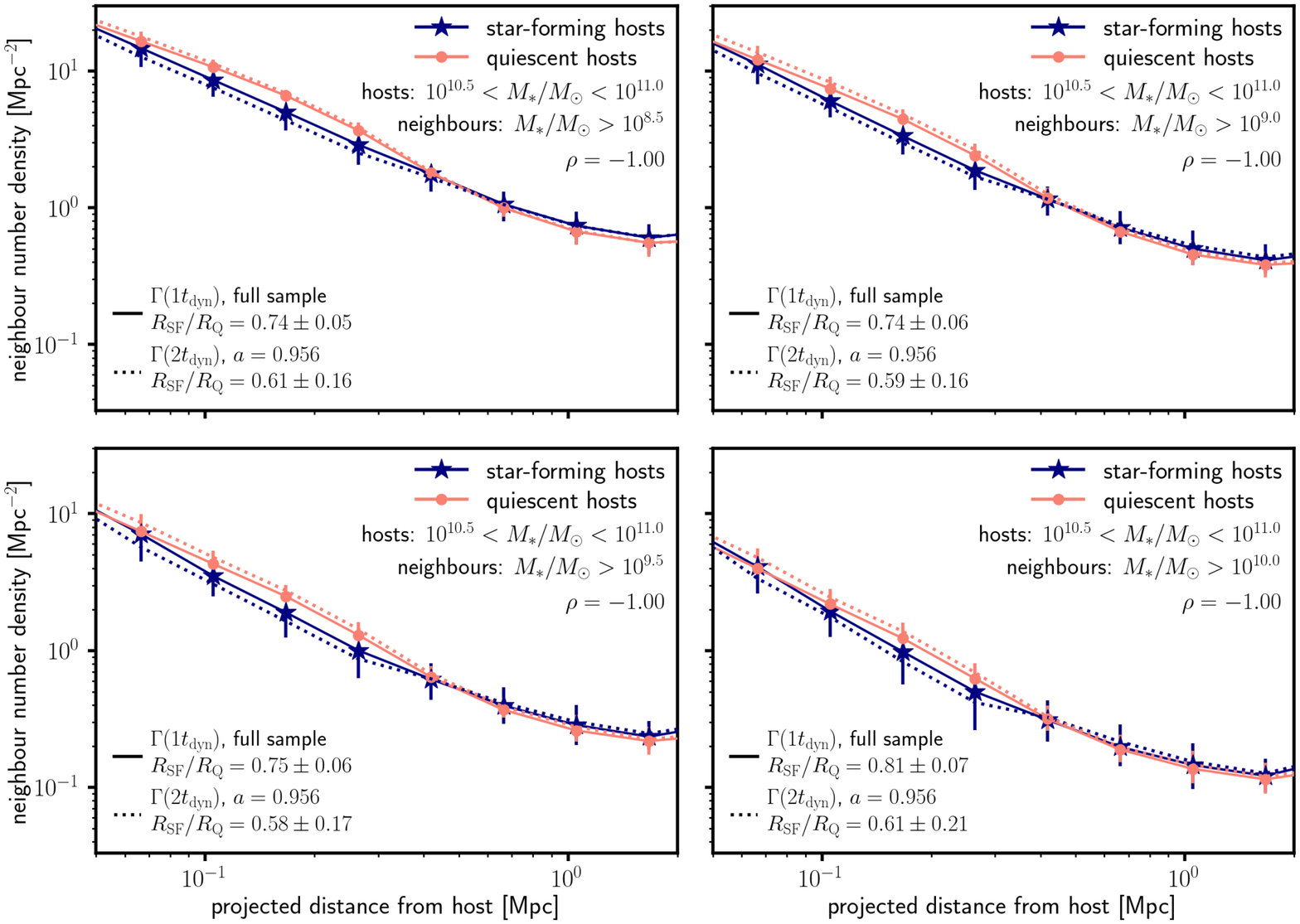}%\\[-2ex]
    \caption{When binning isolated hosts from the \UM{} into star-forming versus quiescent bins based on their specific halo accretion rates over the past $2t_\mathrm{dyn}$ (dotted lines) instead of their past $t_\mathrm{dyn}$ (solid lines), the predicted shape ratios for $\rho=-1.0$ are much smaller than 1 (corresponding to $\rho = 0.0)$. Based on this finding, we would expect to see smaller shape ratios for correlation between dark matter accretion and star formation over longer timescales.}
    \label{fig:2tdyn_rho_-100}
\end{figure*}

To test how strongly our results depend on the definition of specific halo accretion rates (Eq.\ \ref{eq:spec_acc_rate}) as measured over the past $t_\mathrm{dyn}$, we repeated our analysis of neighbour density distributions around isolated hosts with $10.5 < \log_{10}(M_*/M_\odot) < 11.0$ using specific halo accretion rates calculated over the past $2t_\mathrm{dyn}$. In Figs. \ref{fig:2tdyn_rho_0}, \ref{fig:2tdyn_rho_100}, and \ref{fig:2tdyn_rho_-100}, we compare the neighbour density distributions around isolated hosts binned by specific accretion rates over the past $1t_\mathrm{dyn}$ (solid lines) and $2t_\mathrm{dyn}$ (dotted lines) for $\rho=0$, 1, and -1, respectively. We find that the shape ratio deviates farther from 1.0 (the expected shape ratio for no correlation) with longer timescales for the specific halo accretion rates.

\section{Impact of the \UM{} Orphan Model}
\label{app:orphan}

\begin{figure*}
    \centering
    \includegraphics[width=\textwidth]{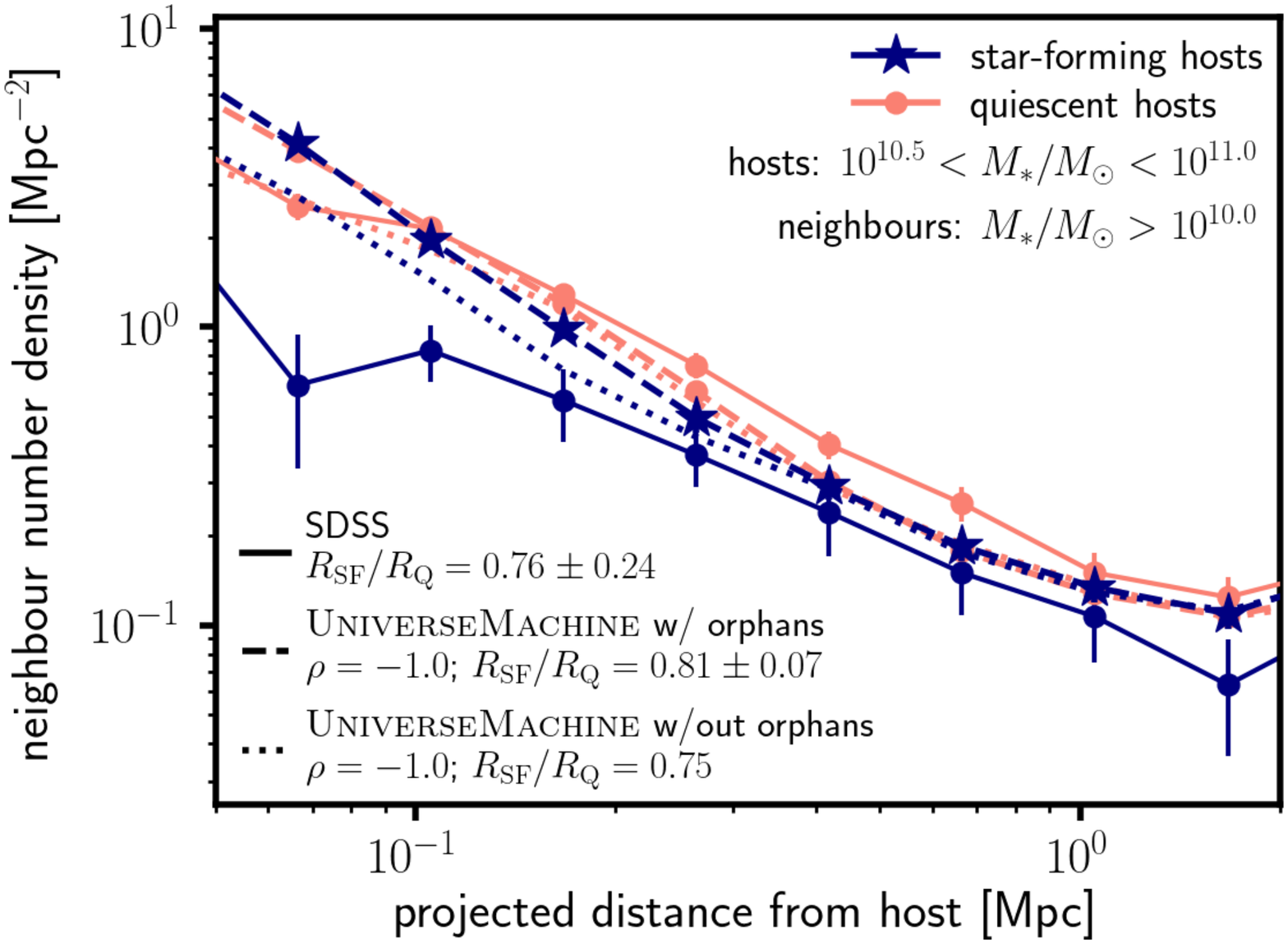}\\[-2ex]
    \caption{Excluding the \UM{} orphan model from the analysis results in a slightly lower density of neighbours close to the host (dotted versus dashed lines), but it does not significantly change the predicted shape ratio. Additionally, we note that this change in the neighbour density distribution does not fully account for the deficit of neighbours observed close to the hosts within the SDSS data.}
    \label{fig:no_orphan}
\end{figure*}

To test the impact of the orphan model in the \UM{}, we replicated our analysis by excluding orphans in the simulation data (Fig.~\ref{fig:no_orphan}). As expected, removing the orphan model slightly reduces the neighbour number density close to the host, but it does not significantly change the predicted shape ratios.

\section{Robustness of the Shape Ratio Metric}
\label{app:shape_ratio}

Our analysis method uses a \textit{shape ratio} parameter $R_\mathrm{SF}/R_\mathrm{Q}$ which compares the shapes of the neighbour density distributions around star-forming and quiescent isolated hosts (Eq.~\ref{eq:shape_ratio} in \S\ref{sec:shape_ratio}). However, as described in \S\ref{sec:density_offset}, we have a systematic offset between observed and simulated neighbour density distributions because of assumed observational biases in stellar masses from the \UM{} catalogues. This offset should only affect the normalisation, but not the shape, of these neighbour density distributions. We test this in both the observational and simulation data by introducing stellar mass offsets (\ref{app:shape_ratio_obs} and \ref{app:shape_ratio_th}, respectively). 

\subsection{Observational Stellar Mass Offset}
\label{app:shape_ratio_obs}

\begin{figure*}
    \centering
    \includegraphics[width=\textwidth]{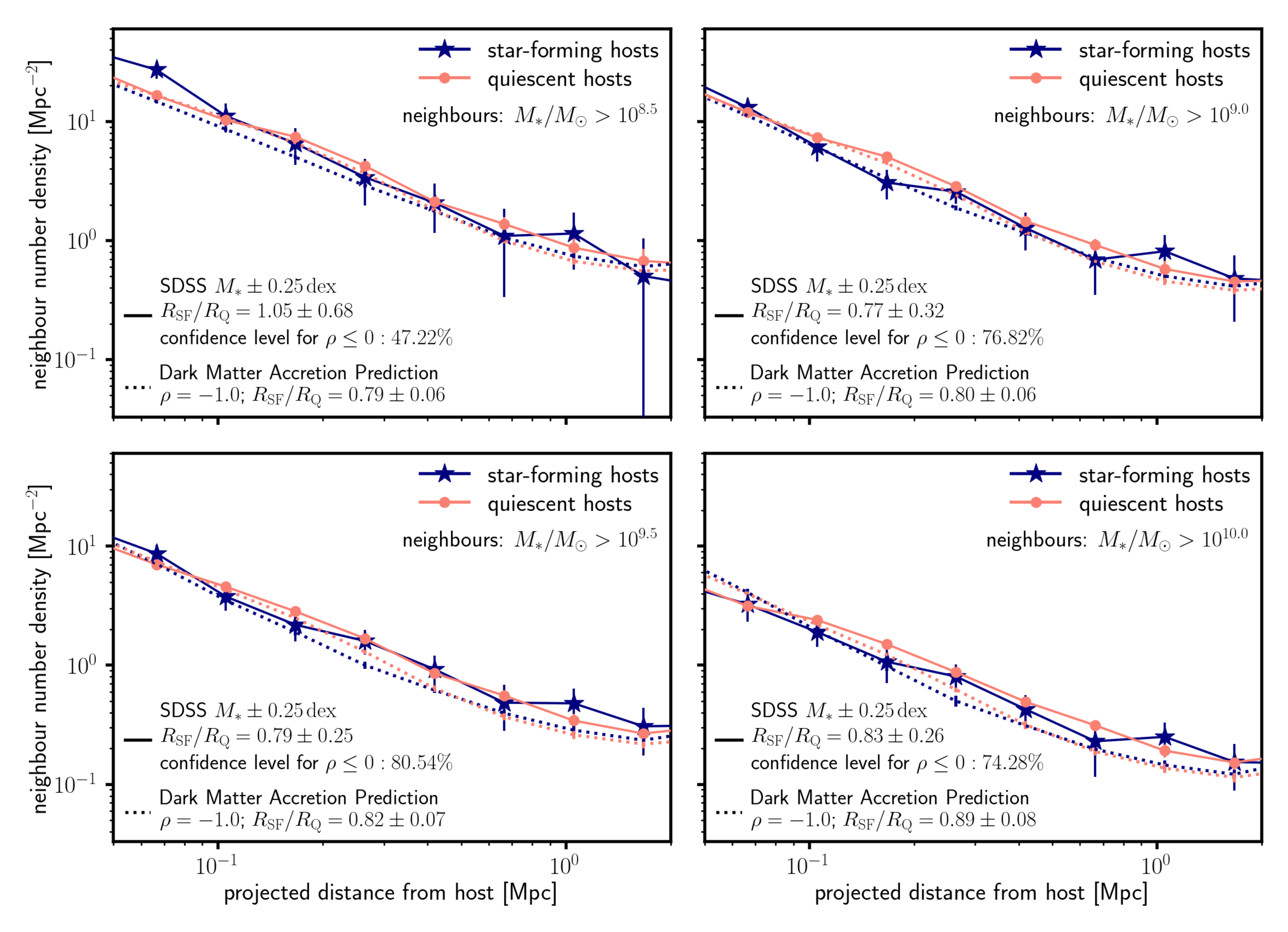}\\[-3ex]
    \caption{By introducing an offset to stellar masses in the SDSS DR16 catalogues (-0.25 dex for star-forming galaxies, +0.25 dex for quiescent galaxies), neighbour density distributions more closely match predictions from simulated \UM{} data, and the resulting shape ratios are still consistent with correlation strengths $\rho \leq 0$ with $\gtrsim 75\%$ confidence. The dark matter accretion predictions are the same neighbour density distributions from the \UM{} plotted in Fig.~\ref{fig:results}.}
    \label{fig:shape_ratio_obstest}
\end{figure*}

Based on Fig.~\ref{fig:results}, the star-forming isolated hosts from the SDSS have a lower normalisation than quiescent isolated hosts. We adjusted the stellar masses of star-forming galaxies in the SDSS ($\mathrm{SSFR} > 10^{-11} \mathrm{yr}^{-1}$) by -0.25 dex and the quiescent galaxies by +0.25 dex. We then identified isolated hosts based on these adjusted stellar masses (i.e., no galaxy with a larger adjusted stellar mass within 2 Mpc projected distance and 1000 km/s velocity distance) and repeated our analysis. However, because the sample size of star-forming hosts dropped significantly, we selected isolated hosts with $10.75 < \log_{10}(M_{*,\mathrm{offset}}/M_\odot) < 11.25$ to increase signal-to-noise. As shown in Fig.~\ref{fig:shape_ratio_obstest}, the resulting neighbour density distributions more closely match the simulated neighbour density distributions from the \UM{}, and the shape ratios are still consistent with $\rho \leq  0$ with $\gtrsim 75\%$ confidence for all selection limits with neighbours with $M_* > 10^{9.0} M_\odot$.

\subsection{Simulated Stellar Mass Offset}
\label{app:shape_ratio_th}

\begin{figure*}
    \centering
    \includegraphics[width=\textwidth]{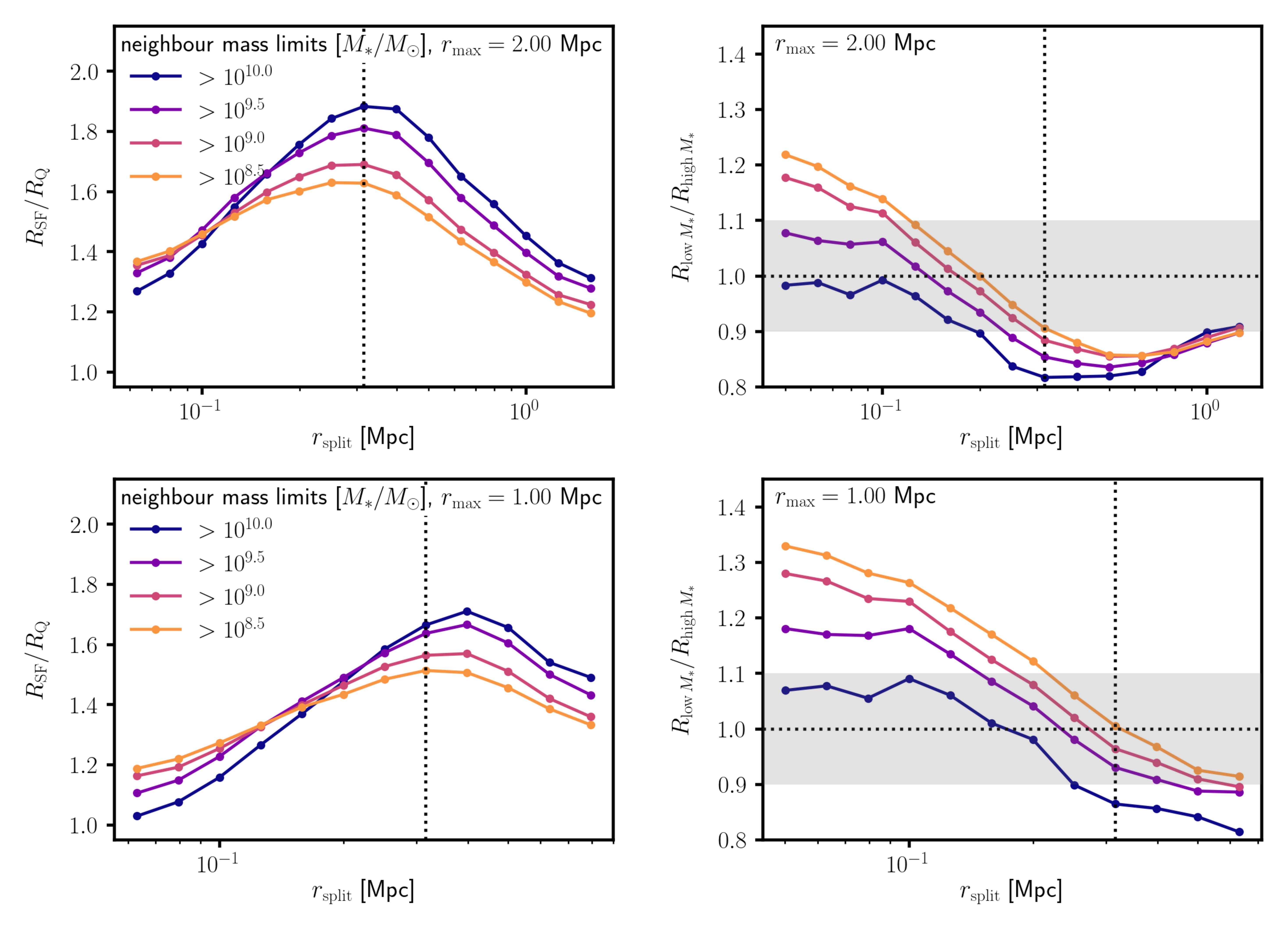}\\[-3ex]
    \caption{Our shape ratio parameter is not significantly affected by stellar mass offsets, and as shown in Table \ref{tab:shape_ratio_modify}, even if we adjust the radial limits to improve the independence of our shape ratio, our results are still consistent with $\rho \leq 0$.
    The two rows show results for different choices of $r_\mathrm{max}$, the outer limit for neighbours included in the shape ratio calculation. The left column replicates Fig.~\ref{fig:shape_ratio}, which plots the shape ratio $R_\mathrm{SF}/R_\mathrm{Q}$ predicted for a correlation strength $\rho=1.0$ for different values of $r_\mathrm{split}$ between the inner and outer regions around isolated hosts with $10.5 < \log_{10}(M_*/M_\odot) < 11.0$ (Eq.~\ref{eq:shape_ratio}). The right column shows the ratio between the shape parameter for the low $M_*$ sample (i.e., isolated hosts in the \UM{} with $10.25 < \log_{10}(M_*/M_\odot) < 10.75$) versus the high $M_*$ sample (i.e., isolated hosts in the \UM{} with $10.5 < \log_{10}(M_*/M_\odot) < 11.0$). The dotted vertical line at 0.316 Mpc represents the value of $r_\mathrm{split}$ used in our analysis. In the right column, the dashed horizontal line and grey band represent $\pm 10\%$ in the value of $R_{\mathrm{low}\, M_*}/R_{\mathrm{high}\, M_*}$.} 
    \label{fig:shape_ratio_th_ratios}
\end{figure*}

We also tested the robustness of the shape ratio by adjusting stellar masses from the \UM{} as was done for \S\ref{sec:results_UMoffset}. To match the normalisations in Fig.~\ref{fig:results}, we compared the neighbour density distributions around all isolated hosts from the \UM{} with stellar masses $10.5 < \log_{10}(M_*/M_\odot) < 11.0$ versus $10.25 < \log_{10}(M_*/M_\odot) < 10.75$ (referred to as the high $M_*$ host sample and low $M_*$ host sample, respectively). Using these samples, we investigated modifying our shape ratios by adjusting $r_\mathrm{split}$ and/or the maximum distance for neighbours far from the host ($r_\mathrm{max}$, which is set to 2.0 Mpc in Eq.~\ref{eq:shape_ratio}). For example, the bottom right panel of Fig.~\ref{fig:shape_ratio_th_ratios} suggests that $r_\mathrm{max} = 1.0$ Mpc and $r_\mathrm{split} = 0.316$ Mpc result in $R_{\mathrm{low}\, M_*}/R_{\mathrm{high}\, M_*}$ being closer to 1.0. When applying these values to the observed data, we still rule out positive correlations with $\gtrsim 80\%$ confidence (Table \ref{tab:shape_ratio_modify}). However, for our analysis in the paper, we choose to keep our original definition of the shape ratio as the method was decided before determining confidence levels, and changing the method \textit{post hoc} would impact the statistical validity of the interpretation of our results.

A final consideration from a stellar mass offset is that it might affect the effectiveness of our isolation criteria. As reported in \S\ref{sec:UM_sample_stats}, $\sim 97\%$ of isolated hosts with $10.5 < \log_{10}(M_*/M_\odot) < 11.0$ were not satellites of larger haloes in the \UM{}. However, if star-forming hosts are reported as having higher stellar masses in the \UM{} than in the real Universe, a galaxy might be misidentified as `isolated' because our isolation criterion relies on the stellar masses of nearby neighbours. To test this concern, we applied a criterion such that a halo with stellar mass $M_*$ would pass if there were no halo with a stellar mass greater than $M_* + 0.25 \mathrm{dex}$ within 2 Mpc projected physical distance and 1000 km/s velocity distance. Of the haloes that passed this test, $\sim 91\%$ were not satellites. We note that a significant increase in the satellite fraction in our `isolated host' sample could affect the shape of neighbour density distributions, resulting in a lower $R_\mathrm{SF}/R_\mathrm{Q}$. However, because our shape ratios were not significantly different when we adjusted the stellar masses in the SDSS (Fig.~\ref{fig:shape_ratio_obstest} in Appendix \ref{app:shape_ratio_obs}), we do not expect that our analysis would be impacted by an increased satellite fraction.

\begin{table}
\begin{center}
\begin{tabular}{c|c|c}
    Neighbour $M_*$ Selection & Modified  & Confidence Level\\
    $\log_{10}(M_*/M_\odot)$ & $R_\mathrm{SF}/R_\mathrm{Q}$ & $\rho \leq 0.0$\\
    \hline 
   8.50 & $0.782 \pm 0.238$ & 81.99\% \\
    9.00 & $0.708 \pm 0.220$ & 90.81\% \\
    9.50 & $0.621 \pm 0.169$ & 98.77\% \\
     10.00 & $0.697 \pm 0.170$ & 96.21\% \\
    \hline
\end{tabular}
\caption{Shape ratios $R_\mathrm{SF}/R_\mathrm{Q}$ for SDSS neighbour density distributions using $r_\mathrm{max}=1.0$ Mpc and $r_\mathrm{split} = 0.316$ Mpc as a test of the robustness of our analysis technique (Fig.~\ref{fig:shape_ratio_th_ratios}). The modified shape ratios are still consistent with correlation strengths $\rho \leq 0$. The increased confidence levels are due to the fact that the observed neighbour density distributions have more noise at larger distances from the isolated hosts, and the smaller $r_\mathrm{max}$ cutoff does not include these annuli. Nonetheless, as explained in Appendix \ref{app:shape_ratio_th}, we do not adopt this different cutoff for the rest of the paper.}
\label{tab:shape_ratio_modify}
\end{center}
\end{table}

\section{Additional Plots of  \texorpdfstring{$\frac{R_\mathrm{SF}}{R_\mathrm{Q}}(\rho)$}{Shape Ratios as a Function of Correlation Strengths}}
\label{app:shape_ratio_plots}

To emphasize our findings, we present additional plots similar to Fig.~\ref{fig:shape_ratio_results_Ms_offset_9.5} that compare the observed shape ratio from the SDSS (grey horizontal band) versus (1) those measured in the \UM{} for the full sample with isolated hosts with  binned into star-forming versus quiescent analogs (\S\ref{sec:UM_accrates} in purple circles and (2) the shape ratio from the \UM{} with isolated hosts with different stellar masses to correspond to the density distribution normalisations from the SDSS (effectively, a test of $\rho=0$) as green squares.

\begin{figure*}
    \centering
    \includegraphics[width=\textwidth]{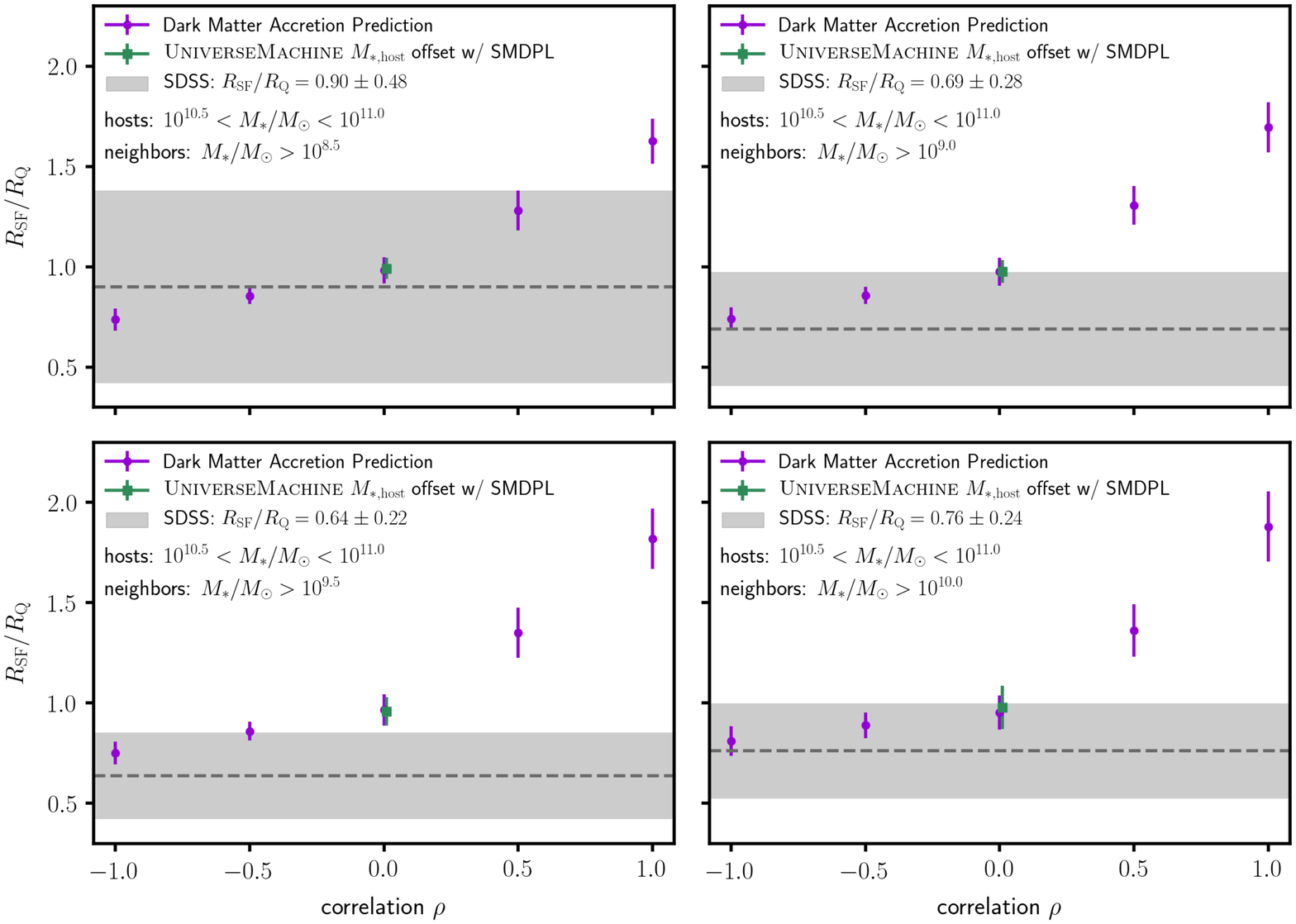}\\[-2ex]
    \caption{Same as Fig.~\ref{fig:shape_ratio_results_Ms_offset_9.5}, but featuring all four neighbour mass selection limits. Regardless of neighbour mass selection, we rule out positive correlations between dark matter accretion and SSFR with $\gtrsim 85\%$ confidence.}
    \label{fig:my_label}
\end{figure*}

\label{lastpage}
\end{document}